\documentclass[aps,floatfix,twocolumn,showpacs,pra]{revtex4}
\usepackage{graphicx,bm}
\usepackage{amsmath}
\usepackage{amssymb}
\usepackage{amsfonts}%
\usepackage{graphicx}
\usepackage{color}

\usepackage[T1]{fontenc}
\usepackage[latin2]{inputenc}
\usepackage{bm}

\newcommand{\bra}[1]{\langle #1|}
\newcommand{\ket}[1]{|#1\rangle}
\newcommand{\meanv}[1]{\left\langle #1 \right\rangle}
\newcommand{\bb}[1]{\left( #1 \right)}

\newcommand{\be}{\begin{equation}}
\newcommand{\ee}{\end{equation}}
\newcommand{\bea}{\begin{eqnarray}}
\newcommand{\eea}{\end{eqnarray}}

\begin{document}
\title{Spin squeezing and EPR entanglement of two bimodal condensates in state-dependent potentials}
\author{Hadrien Kurkjian}
\affiliation{Laboratoire Kastler Brossel,
Ecole Normale Sup\'erieure, UPMC and CNRS,
24 rue Lhomond, 75231 Paris Cedex 05, France}
\author{Krzysztof Paw\l owski}
\affiliation{Laboratoire Kastler Brossel,
Ecole Normale Sup\'erieure, UPMC and CNRS,
24 rue Lhomond, 75231 Paris Cedex 05, France}
\author{Philipp Treutlein}
\affiliation{Department of Physics,
University of Basel, Klingelbergstrasse 82, CH-4056 Basel}
\author{Alice Sinatra}
\affiliation{Laboratoire Kastler Brossel,
Ecole Normale Sup\'erieure, UPMC and CNRS,
24 rue Lhomond, 75231 Paris Cedex 05, France}

\begin{abstract}
We propose and analyze a scheme to entangle the collective spin states of two spatially separated bimodal Bose-Einstein condensates. Using a four-mode approximation for the atomic field, we show that elastic collisions in a state-dependent potential simultaneously create spin-squeezing in each condensate and entangle the collective spins of the two condensates. We investigate mostly analytically the non-local quantum correlations  that arise in this system at short times and show that Einstein-Podolsky-Rosen (EPR) entanglement is generated between the condensates. At long times we point out macroscopic entangled states and explain their structure. The scheme can be implemented with condensates in state-dependent microwave potentials on an atom chip.
\end{abstract}


\maketitle

\section{Introduction}
The internal state of multi-component Bose-Einstein condensates offers {intriguing} possibilities for the creation and investigation of many-particle entanglement. Recent experiments have reported the creation of spin-squeezed states \cite{Treutlein:2010,Oberthaler:2010,Hamley2012}, continuous-variable entanglement \cite{Gross2011}, and twin-atom states \cite{Lucke2011,Bucker:2011}. In these experiments, the entanglement concerns the atoms in a single cloud, and is revealed in collective measurements on the entire system. Alternatively, experiments have explored spin-squeezing with single-component condensates in a double-well potential \cite{Esteve2008,Reichel:2010}, which can be mapped to the two-component internal-state case.
Here we consider a different situation where entanglement is created between the internal states of two spatially separate and individually addressable two-component Bose-Einstein condensates. This system offers the possibility to perform local manipulations and measurements on each of the two spatially separate subsystems, and to study the nonlocal quantum correlations between them.

To describe the interaction-based entangling scheme, we represent each bimodal condensate, labelled by $a$ and $b$,
by a collective spin that is sum of the effective spins 1/2 representing the internal degrees of freedom of each atom.

In a single two-component condensate, the entanglement between the internal degrees of freedom of different atoms, created by the interactions, results in 
quantum correlations between two non commuting components of the collective spin. As a consequence, the fluctuations  of a well chosen linear combinations of these two components are reduced below the quantum limit for independent atoms
\cite{Ueda:1993,Sorensen:2001,YunLi:2009}.

In the generalization that we propose, the entanglement between atoms belonging to different ensembles, results in additional correlations between the collective spins $\hat{\vec{S}}_a$ and $\hat{\vec{S}}_b$. Fluctuations in a well chosen quadrature, linear combination of spin $a$ and $b$ components, are then reduced, and informations about the spin in system $b$ can be inferred from a measurement in system $a$. We show furthermore that these correlations are non-classical  and non-local in the sense of Einstein-Podolsky-Rosen (EPR) \cite{EPR:1935}.
As a physical realization we propose to use controlled interactions in state-dependent potentials, for example in an optical trap or in a microwave trap {on an atom chip}. 

Other strategies to create EPR-type entanglement between two Bose condensates have been proposed in \cite{He:2011,BarGill:2011}.
We also note that in room-temperature atomic vapor cells containing $\simeq 10^{12}$ atoms, entanglement between two collective spins  has been successfully created experimentally using the interaction of the atoms with a pulse of light followed by a measurement \cite{Julsgaard2001}. 
The Bose-condensed system we consider here contains a few tens to a few thousand atoms. It offers exceptional coherent control and measurement down to the single-atom level giving access to the discrete variables regime. Moreover, the fact that the system is almost isolated in principle allows for creation of highly entangled quantum states.

We explain the scheme we have in mind in Sec. \ref{sec:setup} and the theoretical model in Sec. \ref{sec:model}.
Section \ref{sec:long_times} is devoted to long times, when macroscopic superpositions entangled between the two ensembles occur, in correspondence with sharp dips in the entropy of entanglement. In the last Section \ref{sec:short_times} we concentrate on short-times EPR-like quantum correlations. To highlight the non-classical nature of these correlations, we use a criterion introduced in \cite{Wiseman2007,Opanchuk2012} which involves the covariance matrix of the collective spins components. Non-classical correlations can be obtained for a broad range of experimental parameters.

\section{Considered scheme}
\label{sec:setup}
\begin{figure}[t]
\centering
\includegraphics[width=0.3\textwidth]{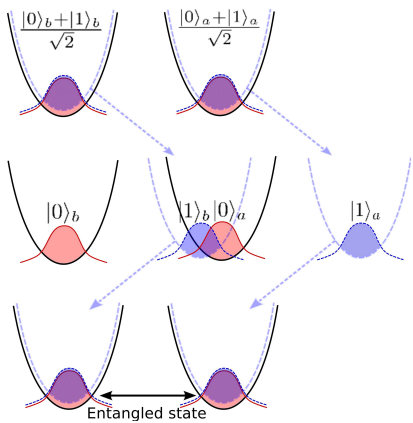}
\caption{(Color online) Sequence allowing to entangle condensate {$a$ (right well) and $b$ (left well)} via controlled {collisional} interaction in state dependent trapping potentials. The interaction phase, where the squeezing and the correlations are created, 
is depicted in the central panel.
\label{fig:setup}}
\end{figure}

The experimental situation we consider is depicted in Fig.~\ref{fig:setup}. It involves two Bose-Einstein condensates $a$ and $b$ in two spatially separated {potential} wells. Initially the condensates are in the ground state $\ket{0}$ of some internal transition $\ket{0}-\ket{1}$. In the case of ${}^{87}$Rb atoms
one can take for example $\ket{0}=\ket{F=1,m_F=-1}$ and $\ket{1}=\ket{F=2,m_F=1}$. At time $t=0$ an electromagnetic pulse prepares each atom in an {equal} superposition of internal states $\ket{0}$ and $\ket{1}$. {After this} we imagine that, while the traps for state
$\ket{0}$ do not move, the traps for state $\ket{1}$ are moved out so that the component $\ket{1}$ of {condensate $b$} interacts with the component $\ket{0}$ of {condensate $a$}. After a given interaction time, the traps for state $\ket{1}$ 
come back to their initial position. Due to atomic interaction within each component $\ket{0}$ and $\ket{1}$ the two condensates $a$ and $b$ are spin-squeezed. Moreover, due to the crossed interaction between {$\ket{1}_b$ and $\ket{0}_a$}, the two condensates are entangled. This scheme is a direct generalization of the ``collisional gate'' scheme proposed in {\cite{Zoller:2000,Treutlein2006} for two individual atoms. It could be implemented using optical potentials as in \cite{Zoller:2000,Bloch:2003}, or with microwave potentials on an atom chip \cite{Boehi:2009,Treutlein:2010}.}

\section{Four-mode model}
\label{sec:model}
\subsection{Hamiltonian}

{Collisional} interactions between cold atoms occur only  in $s$-wave and can be modeled by a zero-range potential (see e.g.\cite{LesHouches99}). This leads to the interaction Hamiltonian
\begin{multline}
 \hat{H}_{\rm int} =   g_{01}\int \hat{\Psi}_{0}^{\dagger}\hat{\Psi}_{0}\hat{\Psi}_{1}^{\dagger}\hat{\Psi}_{1}  + \sum_{\epsilon \in \{0,1\}}\frac{g_{\epsilon \epsilon} }{2}\int \hat{\Psi}_{\epsilon}^{\dagger}\hat{\Psi}_{\epsilon}^{\dagger}\hat{\Psi}_{\epsilon}\hat{\Psi}_{\epsilon},
\end{multline}
where $\hat{\Psi}_{0(1)}$ is the field operator for atoms in internal state $0(1)$  and 
$g_{\epsilon \epsilon'}$ is the coupling constant for contact interactions between atoms in states $\epsilon$ and $\epsilon'$, related to the s-wave scattering length $a_{\epsilon \epsilon'}$ of the interaction potential by $g_{\epsilon \epsilon'} =4\pi \hbar^2 a_{\epsilon \epsilon'} / m$.
We now expand each of the field operators  $\hat{\Psi}_0$ and $\hat{\Psi}_1$ over the two spatial modes that will be macroscopically populated:
\begin{eqnarray}
 \hat{\Psi}_0 &=& \phi_{0, a} \hat{a}_0 + \phi_{0, b} \hat{b}_0, \\
 \hat{\Psi}_1 &=& \phi_{1, a} \hat{a}_1 + \phi_{1, b} \hat{b}_1.
\end{eqnarray}
Just after the coupling pulse one has $\phi_{1, a}=\phi_{0, a}$ and $\phi_{1, b}=\phi_{0, b}$.
In the scheme we consider, during the interaction time (central panel of Fig.~\ref{fig:setup}), only {$\phi_{1, b}$ and $\phi_{0, a}$} have a non-zero overlap. 

{For the two condensates $a$ and $b$ we introduce the collective spin operators  $\hat{\vec{S}}^a$
and  $\hat{\vec{S}}^b$, respectively.} We have for example for $a$
\begin{eqnarray}
\hat{S}_x^a&=&\bb{\hat{a}_1^{\dagger} \hat{a}_0 + \hat{a}_1^{\dagger} \hat{a}_0}/2, \notag \\ 
\hat{S}_y^a&=&\bb{\hat{a}_1^{\dagger} \hat{a}_0 - \hat{a}_0^{\dagger} \hat{a}_1}/{2 i}, \notag \\
\hat{S}_z^{a} &=&  \bb{\hat{a}_1^{\dagger} \hat{a}_1 - \hat{a}_0^{\dagger} \hat{a}_0}/2,  
\end{eqnarray}
and similarly for $b$. The non-linearities leading to squeezing and entanglement will be ruled by the 
parameters $\chi_{\epsilon,\sigma}$ and $\chi_{ab}$ that depend on the interaction constants, on the modal wave functions and their overlap:
\begin{eqnarray}
\hbar\chi_{\epsilon,\sigma} &=& \frac{g_{\epsilon \epsilon}}{2}\int |\phi_{\epsilon,\sigma}|^4, \quad \quad  \epsilon=0,1, \quad \sigma=a,b, \label{eq:def_chi_epsilon_sigma}\\
\hbar\chi_{ab} & = & g_{01}\int |\phi_{0,a}|^2 |\phi_{1,b}|^2.
\end{eqnarray}
With these notations and for a system with initially $N$ atoms in each well, one can rewrite the interaction Hamiltonian as
\begin{multline}
\label{eq:Hamiltonian}
 \hat{H}_{\rm int}/\hbar  = 
{\sum_{\epsilon =0, 1}} \chi_{\epsilon,a} \bb{\hat{S}_z^{a}}^2 + {\sum_{\epsilon =0, 1}} \chi_{\epsilon,b} \bb{\hat{S}_z^{b}}^2
 -\chi_{ab} \hat{S}_z^{a} \hat{S}_z^{b}  \\
 + \frac{\chi_{ab}N}{2}\bb{\hat{S}_z^b - \hat{S}_z^a} +\sum_{\sigma = a, b} N \hat{S}_z^{\sigma} (\chi_{1,\sigma} - \chi_{0,\sigma})  ,
\end{multline}
where we have omitted a constant term.
The first and second term generate squeezing within the condensate $a$ and $b$, respectively. The third term is the one responsible for the $a-b$ entanglement. The linear terms in the second line (that give a clock shift due to interactions in each condensate) can in principle be removed with a $\pi$-pulse that reverses all the spins at half evolution time and we shall neglect them in the following.
In the particular case where $g_{00}=g_{11}$ and all the wave functions have the same shape, we set 
\be
\label{eq:def_chi}
{\chi_{\epsilon,\sigma}}\equiv\chi/2 \quad \quad \forall \epsilon=0,1 \quad \forall \sigma=a,b
\ee
and simplify the non-linear part of the Hamiltonian to
\begin{equation}
\label{eq:HamiltonianSimple}
 \hat{H}_{\rm int}^{\rm nl}/\hbar  = \chi \bb{\hat{S}_z^{a}}^2+ \chi \bb{\hat{S}_z^{b}}^2  -\chi_{ab} \hat{S}_z^{a} \hat{S}_z^{b}. 
 \end{equation}

\subsection{State evolution}
To compute the evolution of the state under this Hamiltonian, we expand it in the Fock basis. 
Immediately after the pulse, the state is a product of two phase states  
{$\ket{\Psi (0)}= \ket{\phi_a=0}_{\rm ph} \ket{\phi_b=0}_{\rm ph}$}
where
\be
\ket{\phi_a=\phi}_{\rm ph} \equiv \frac{1}{\sqrt{2^NN!}} \left( e^{-i \phi/2 }a_0^\dagger + e^{i \phi/2 }a_1^\dagger  \right)^N |0\rangle.
\label{eq:def_phase_state} 
\ee
We expand the phase states over Fock states
\begin{eqnarray}
\label{eq:phasestate} \ket{\Psi (0)} &=& \frac{1}{2^{N}N!}\bb{\hat{a}_0^\dagger + \hat{a}_1^\dagger}^{N}\bb{\hat{b}_0^\dagger + \hat{b}_1^\dagger}^{N}\ket{0}\\
 &=& \frac{1}{2^{N}}\sum_{n_a, n_b} \sqrt{\binom{N}{n_a} \binom{N}{n_b} } \ket{n_a, n_b}_{\rm F} \notag
\end{eqnarray}
where we have introduced the notation
\be
\label{eq:notation_Fock}
\ket{n_a, n_b}_{\rm F} \equiv \frac{(\hat{a}_1^\dagger)^{n_a} (\hat{a}_0^\dagger)^{N-n_a} (\hat{b}_1^\dagger)^{n_b} (\hat{b}_0^\dagger)^{N-n_b}}{\sqrt{n_a! \, (N-n_a)! \, n_b! \, (N-n_b)! }} \ket{0} .
\ee

The Hamiltonian \eqref{eq:HamiltonianSimple} is diagonal in the Fock basis so that during evolution each Fock state simply acquires a phase factor $\varphi(n_a,n_b)$:
\be
\label{eq:evolvedstate}
 \ket{\Psi (t)} = \frac{1}{2^{N}}\sum_{n_a, n_b} \sqrt{\binom{N}{n_a} \binom{N}{n_b} } e^{-it\varphi(n_a,n_b)} \ket{n_a, n_b}_{\rm F}.
\ee
Introducing the eigenvalues of $\hat{S}_z^{a(b)}$, $\delta n_{a(b)} \equiv n_{a(b)}-N/2$, we can write the phase factor 
$\varphi(n_a,n_b)$ as
\be
\varphi(n_a,n_b)=-\chi_{ab} \delta n_a \delta n_b + \chi \bb{\delta n_a^2 +  \delta n_b^2}.
\ee

\section{Entropy of entanglement and conditional macroscopic superpositions}
\label{sec:long_times}
\begin{figure}
   \begin{minipage}[c]{.49\linewidth}
     \includegraphics[width=\textwidth]{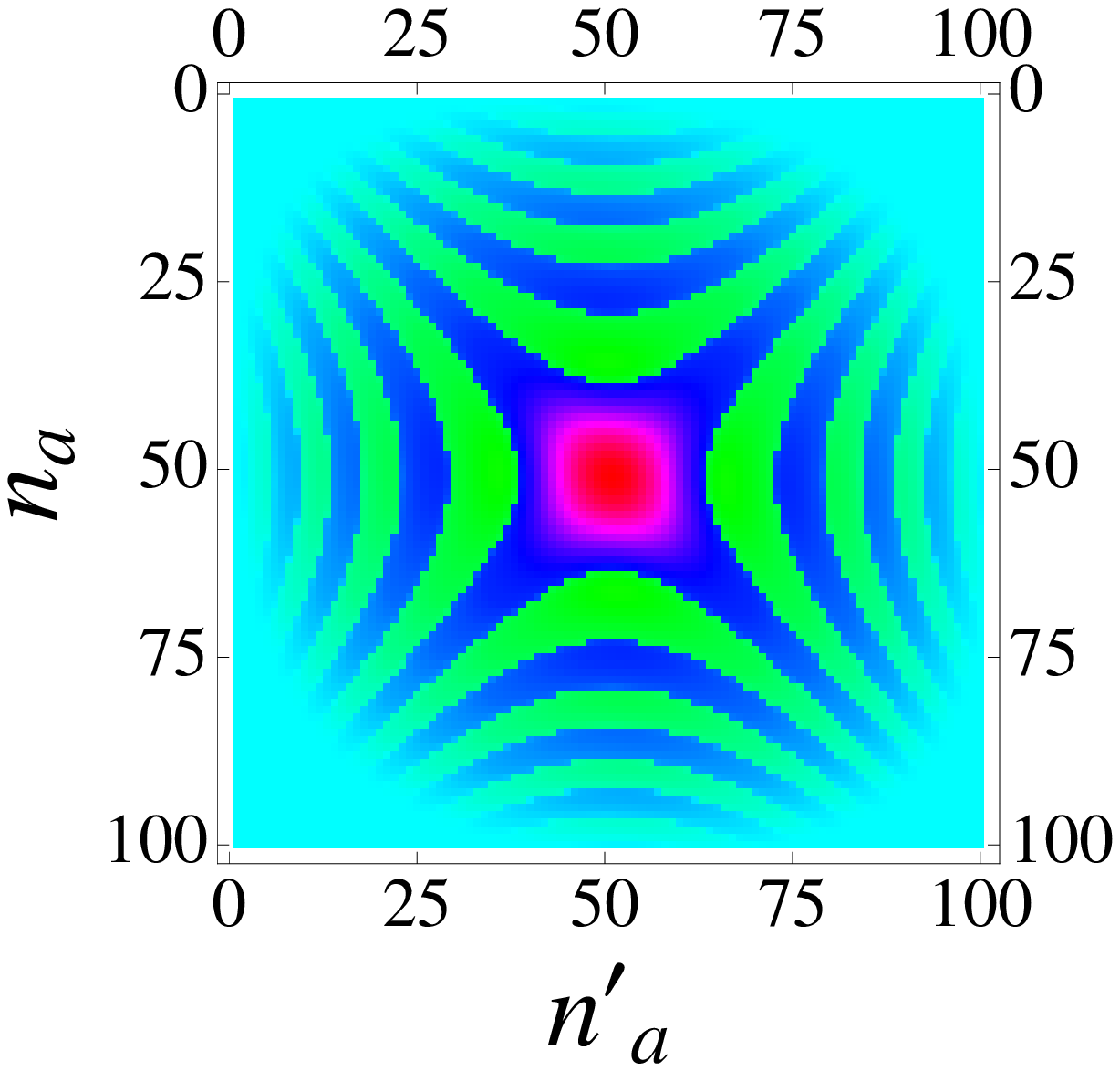}
      \center{$\chi_{ab}t=0.01$}
   \end{minipage} \hfill
   \begin{minipage}[c]{.49\linewidth}
       \includegraphics[width=\textwidth]{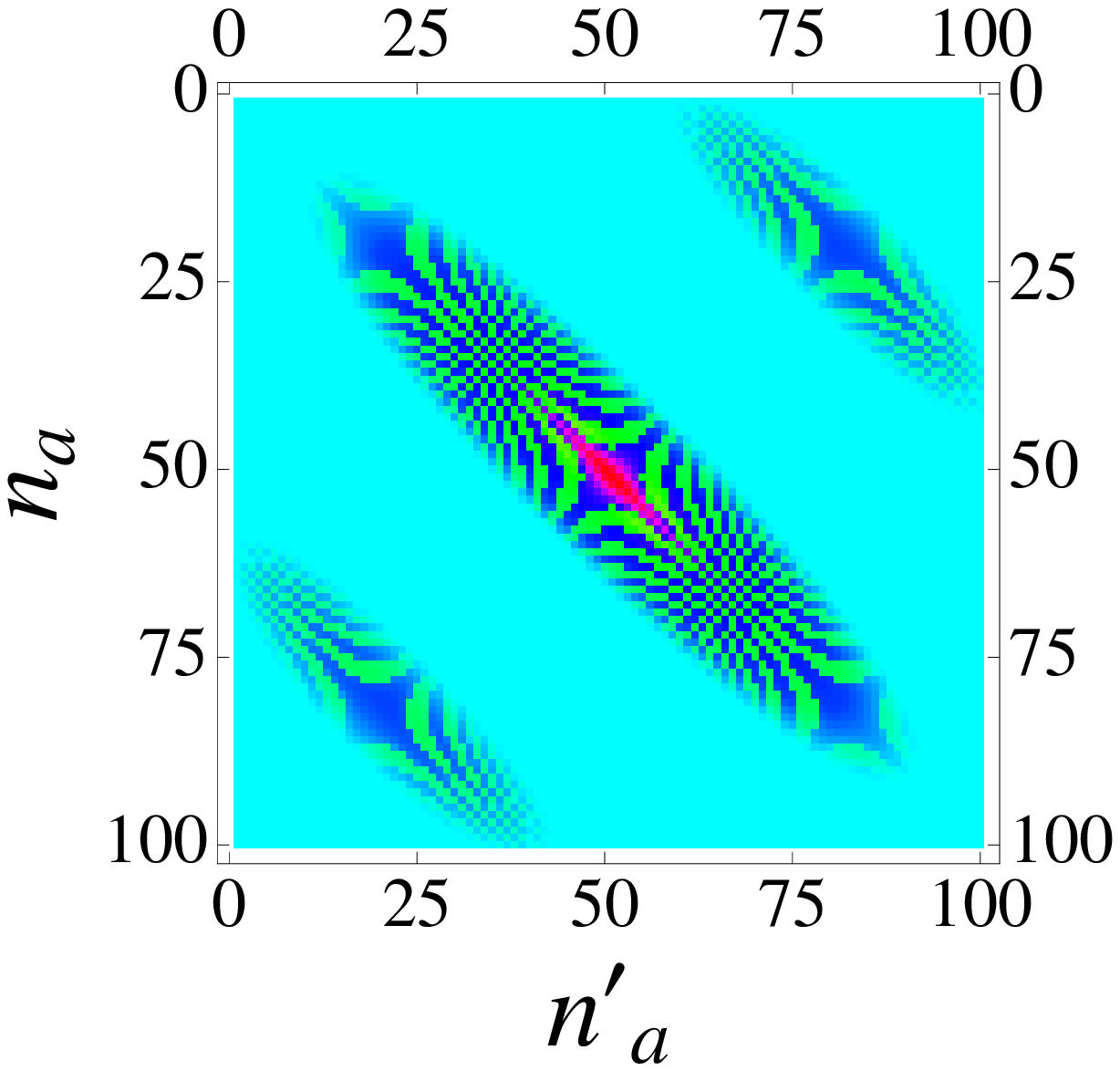}
      \center{ $\chi_{ab}t=0.1$}
   \end{minipage} \\ 
  
   \vspace{0.5cm}
      \begin{minipage}[c]{.49\linewidth}
       \includegraphics[width=\textwidth]{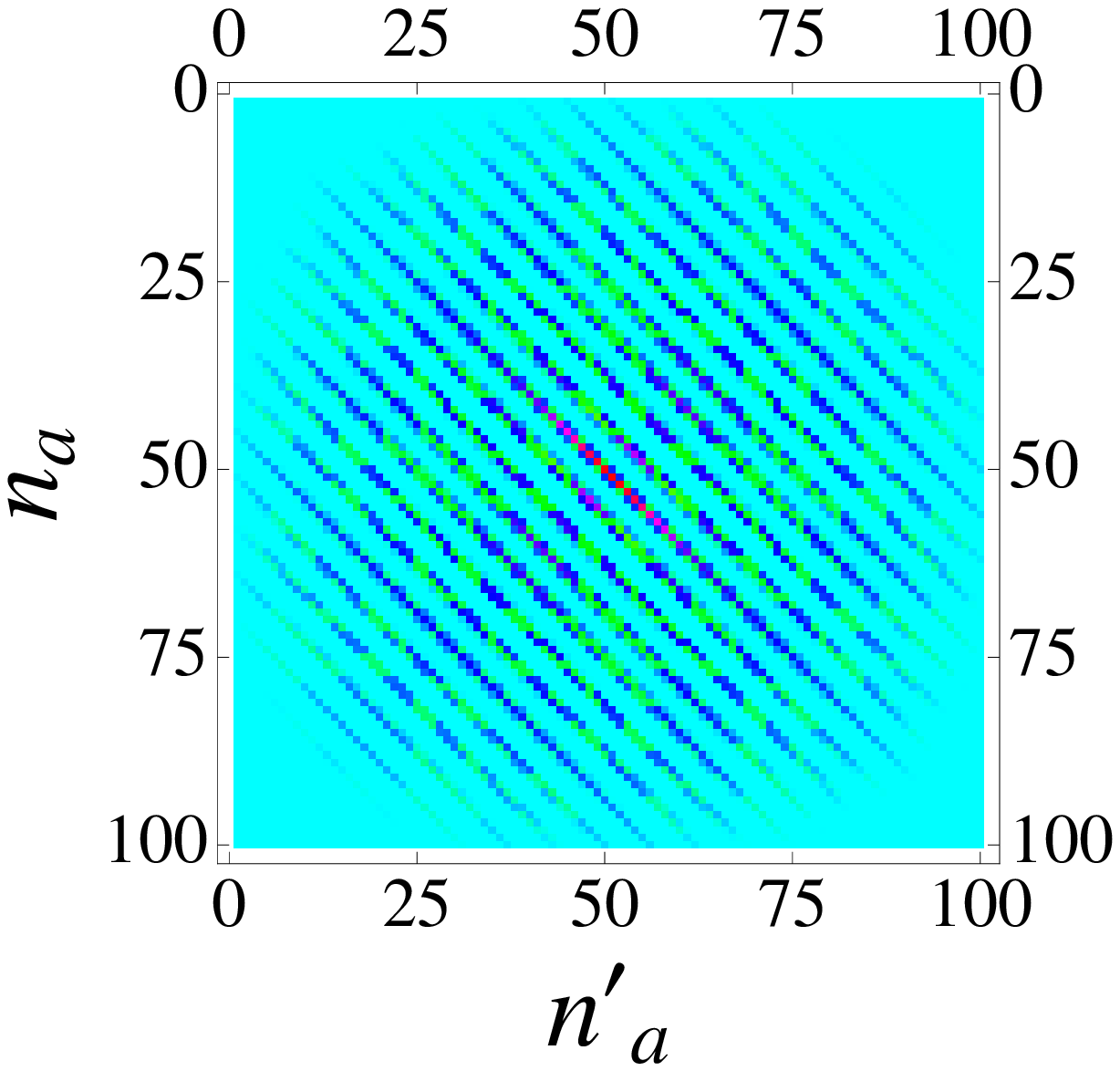}
      \center{ $\chi_{ab}t=1$}
        \end{minipage} \hfill 
      \begin{minipage}[c]{.49\linewidth}
            \includegraphics[width=\textwidth]{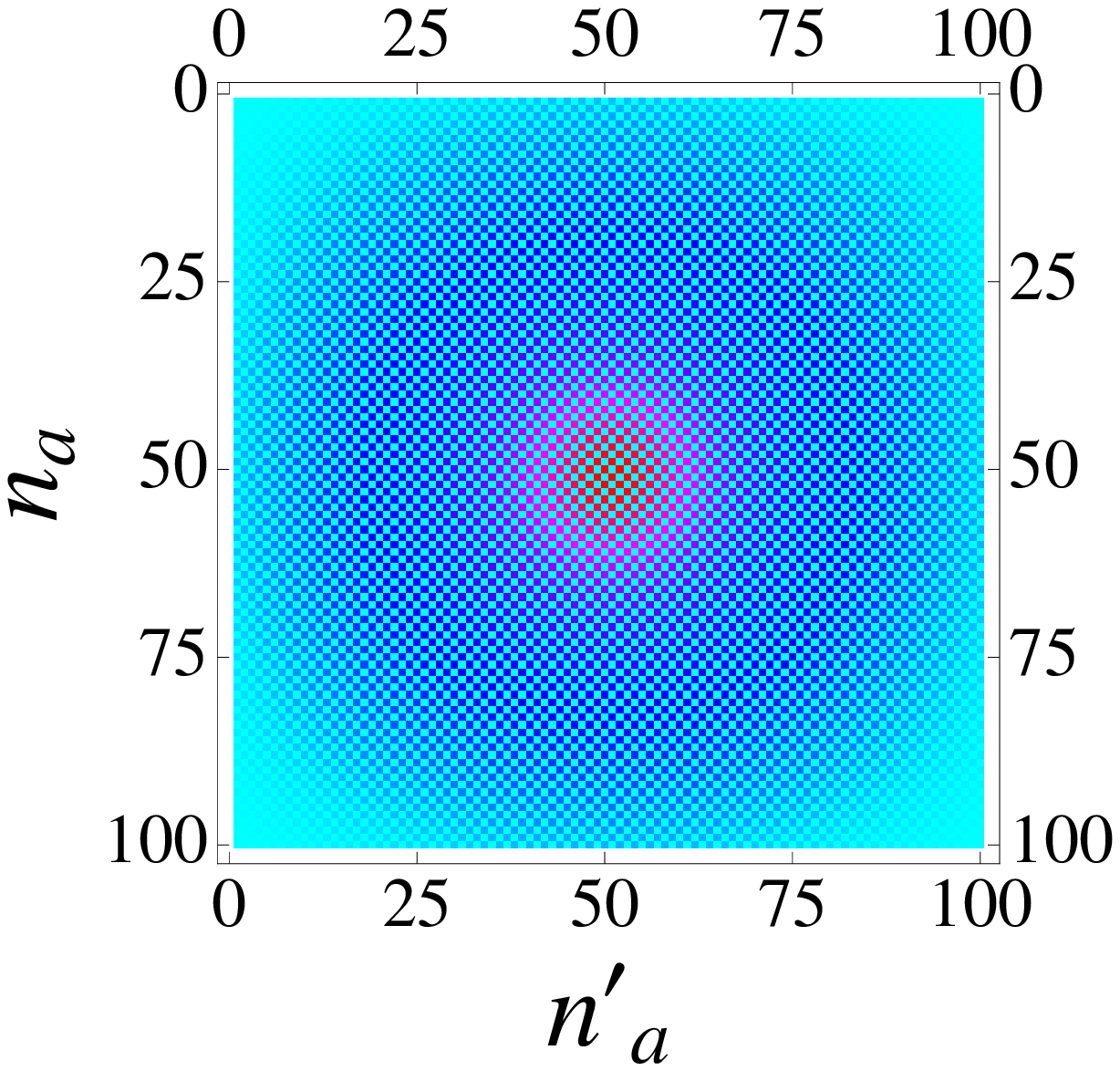}
      \center{$\chi_{ab}t=\pi$}
   \end{minipage}
   \caption{(Color online) Real part of the reduced density matrix in the Fock basis $(\rho_a)_{n_a,n_a'}$ at different times (in units of $1/\chi_{ab}$) for $\chi_{ab}=\chi$ and $N=100$ atoms in each of the two ensembles. 
We start with the density matrix of a phase state (\ref{eq:phasestate}). 
The entanglement dynamics damps out the off-diagonal coefficients after which striped patterns appear. The interpretation of these patterns is given in the text. The color scale ranges from red (largest positive), green, light blue (zero) to dark blue (negative). 
 \label{fig:matricedensite}
}
\end{figure}

To quantify the entanglement between $a$ and $b$ condensates, we use the entropy of entanglement that, for the pure state we consider here, is simply the Von Neumann entropy of the reduced density matrix of $a$ or $b$,
\bea
S&=&-\text{Tr}(\rho_a \log \rho_a), \\
\label{eqn:entropy}
\rho_a&=&\text{Tr}_b \ket{\Psi}\bra{\Psi}.
\label{eqn:rhoB}
\eea
Using the expression (\ref{eq:evolvedstate}) for the state, one can work out the {elements} of the reduced density matrix of system $a$,
\begin{multline}
(\rho_a)_{n_a,n_a'}=\frac{1}{2^N}\sqrt{\binom{N}{n_a} \binom{N}{n_a'}} \exp\bb{-i t \chi(\delta n_a^2-\delta n_a'^2)}\\ \times {\cos^{N}\bb{ \frac{\chi_{ab}  t}{2} (n_a - n_a')}}.
\end{multline}
In Fig.~\ref{fig:matricedensite} we show the evolution of the reduced density matrix in the Fock basis for the particular case where $\chi_{ab}/\chi=1$, corresponding to equal coupling constants $g_{01}=g_{00}=g_{11}$ and perfect overlap of {$\phi_{1,b}$ and $\phi_{0,a}$}.
At long times larger than $1/\sqrt{N}\chi_{ab}$ a striped pattern appears. According to the analysis in \cite{ferrini2010a}, referring to a single BEC,
such structure is characteristic of a mixture of different phase states, the distance between the stripes being inversely proportional to the number of phase states involved. In the extreme case of the lower right panel at time 
\be
t_{\pi}\equiv\pi/\chi_{ab}
\label{eq:tpi}
\ee
we note a  remarkable checkerboard structure. This is a particular striped pattern suggesting that the state of condensate $a$ is a mixture of only two phase states differing by $\pi$ (see Fig. 3 in \cite{ferrini2010a}). 
As the total state is pure, the appearance of such a mixture in the reduced density matrix indicates that (i) the modes $a$ and $b$ are entangled and (ii) that, if expressed in the phase state basis, 
 the global state may be a relatively simple superposition.

\begin{figure}
   \begin{minipage}[c]{\linewidth}
   \center{$(a) \quad \chi=\chi_{ab}$} \\ \vspace{5pt}
     \includegraphics[width=\textwidth]{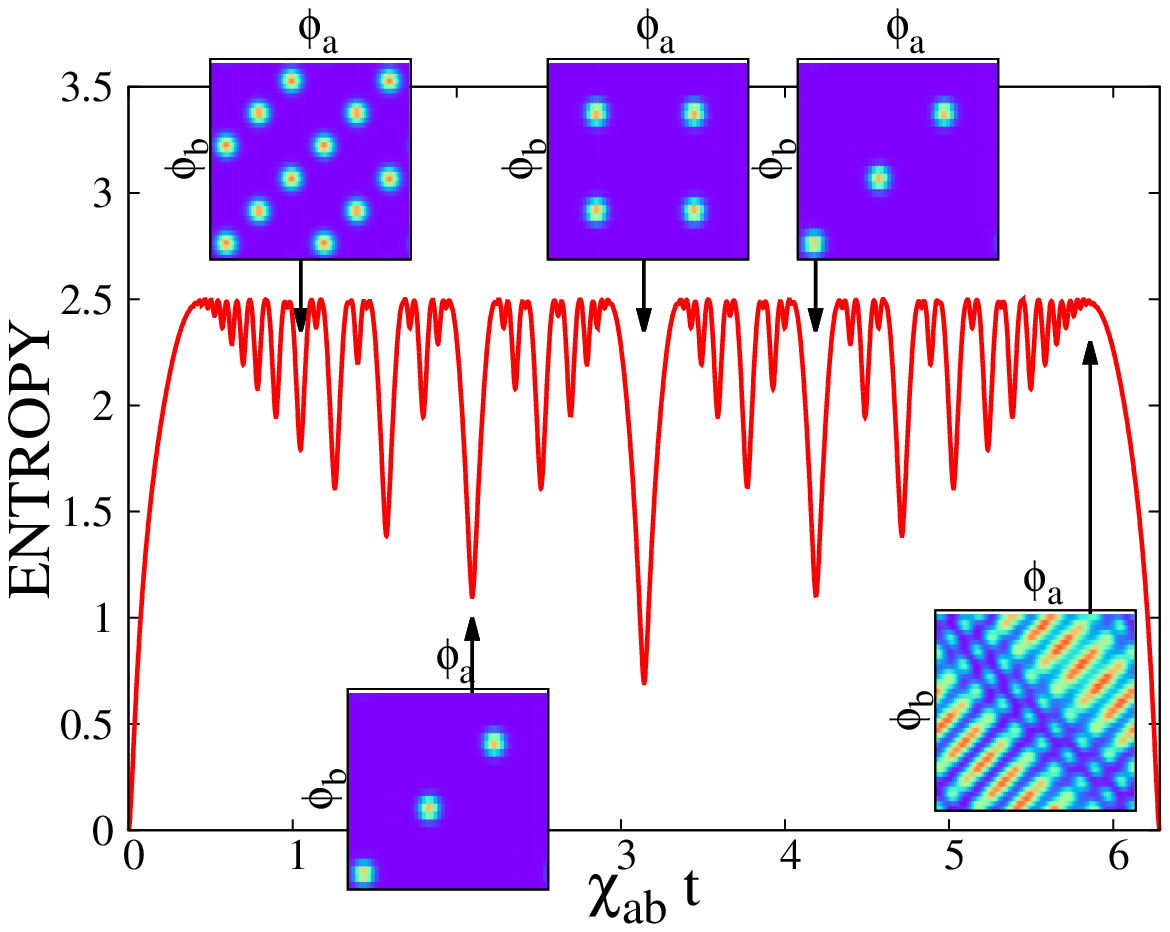}
   \end{minipage} \hfill \\ \vspace{10pt}
   \begin{minipage}[c]{0.49\linewidth}
     \includegraphics[width=\textwidth]{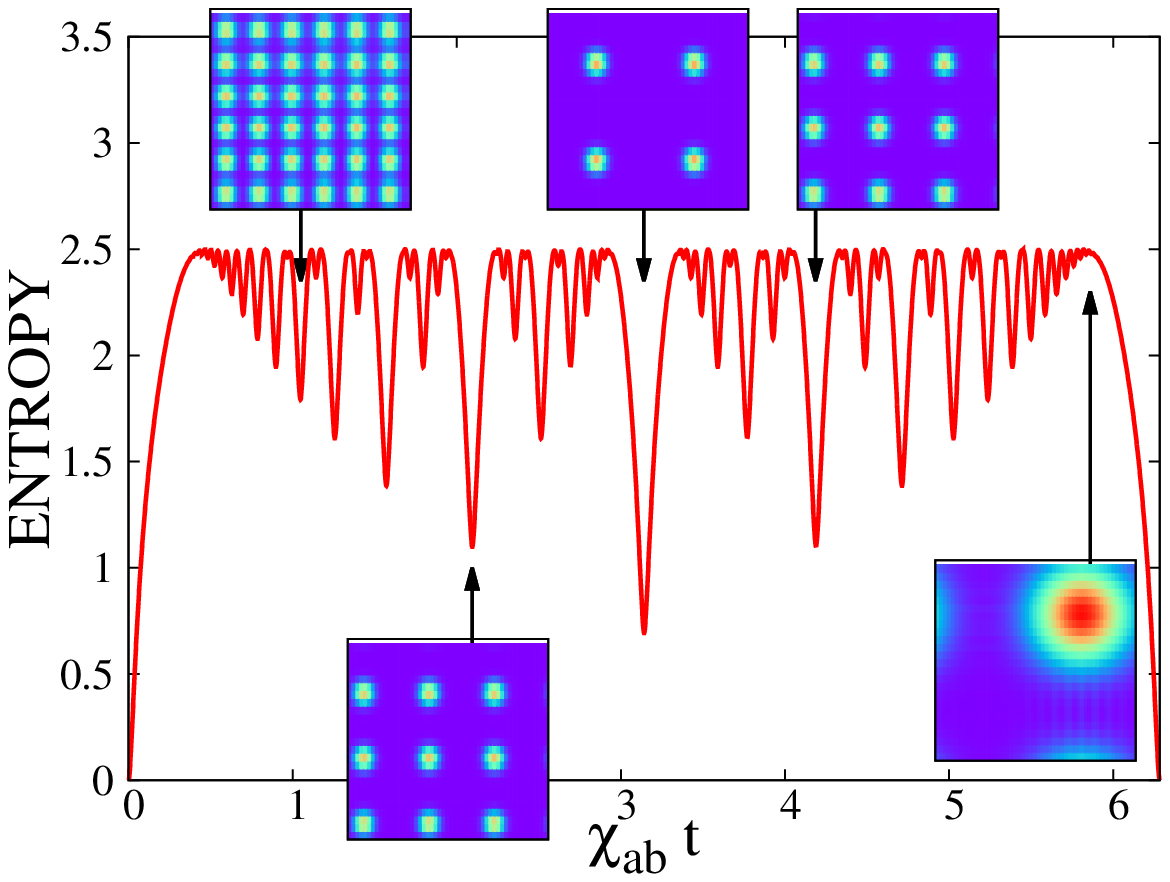}
      \center{$(b) \quad \chi=0$}
      \end{minipage}
       \begin{minipage}[c]{0.49\linewidth}
       \includegraphics[width=\textwidth]{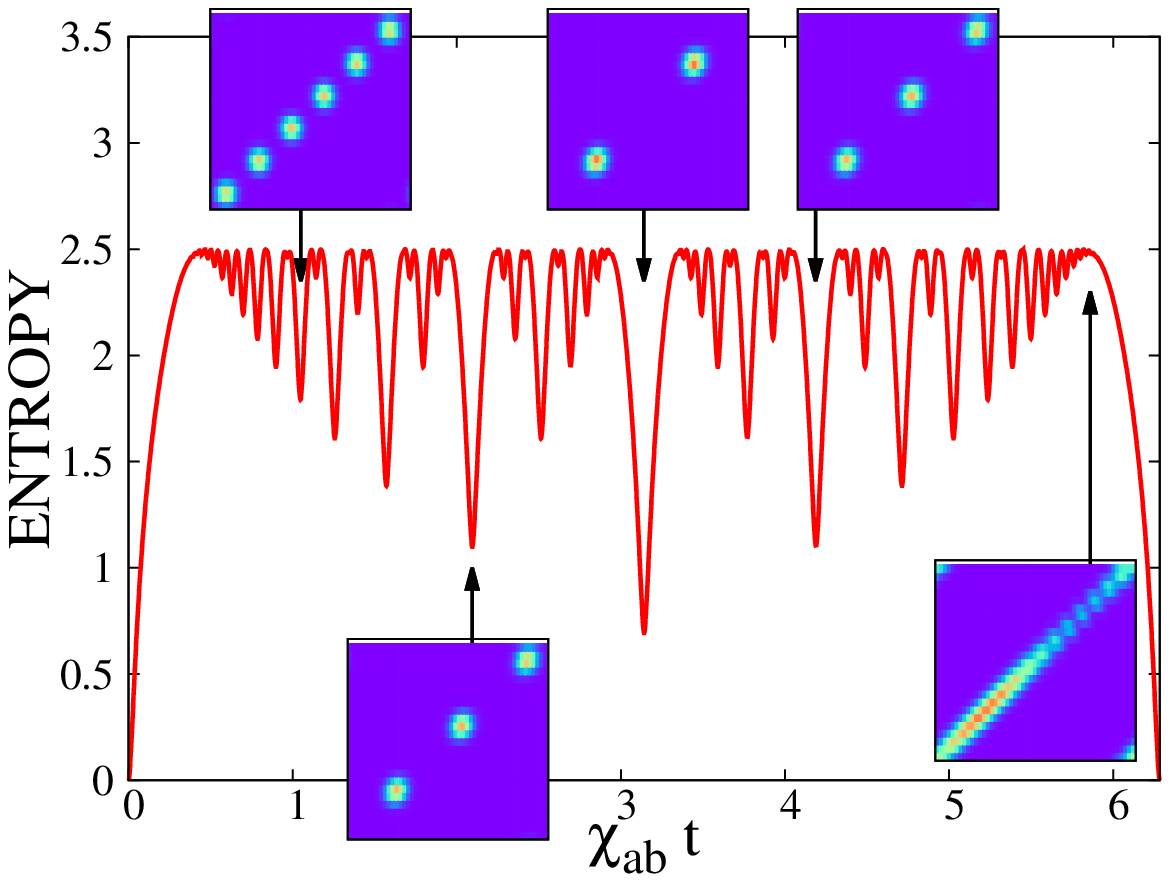}
      \center{ $(c) \quad \chi=0.5\chi_{ab}$}
   \end{minipage} \\ 
\caption{(Color online) Entropy of entanglement between ensembles $a$ and $b$ as a function of time for different values of the ratio $\chi/\chi_{ab}$. 
The number of atoms in $a$ and $b$ is even: $N=50$. 
The insets show the total density matrix in phase state basis { $\rho_{\phi_a,\phi_b}$, see Eq.~(\ref{eq:densitymatrixphase}), } at times 
$ \chi_{ab}t=\pi/3$, $2\pi/3$, $\pi$, $4\pi/3$ and $\bb{2\pi-\frac{1}{2\sqrt{N}}}$. The phases $\phi_a$, $\phi_b$ range in 
$[\pi/2,5\pi/2]$. We note the appearance of simple superpositions of phase states.
\label{fig:entropy}}
\end{figure}

In Fig.~\ref{fig:entropy}a we  plot the entropy of entanglement for $\chi_{ab}/\chi=1$.
Indeed, at time $t_{\pi}=\pi/\chi_{ab}$\, we find $S\simeq \log (2)$, as it should be for a mixture of two almost orthogonal states.
We note that also at {many other} rational fractions of $t_{\pi}$, $t=(2m/q) t_{\pi}$, the entropy of entanglement shows dips
reaching the values $S\simeq\log (q)$.

To understand the structure of the state at these times and the behavior of the entropy it is useful to rewrite the state \eqref{eq:evolvedstate} at time $t$ in the form:
\begin{equation}
 \ket{\Psi(t)}=e^{it\chi_{ab}\hat{S}_z^a\hat{S}_z^b} \ket{\Psi_a,\Psi_b},
\label{eqn:psiT}
\end{equation}
where $\ket{\Psi_a,\Psi_b}$ is the initial state evolved with the squeezing Hamiltonians:
\begin{equation}
\ket{\Psi_a,\Psi_b}=e^{-i\chi t\bb{\bb{\hat{S}_z^a}^2+\bb{\hat{S}_z^b}^2}}\ket{00}_{\rm ph}.
\label{eqn:psiApsiB}
\end{equation}
Here we exploit the shortened notation \mbox{$\ket{\alpha\beta}_{\rm ph}\equiv\ket{\phi_a=\alpha}_{\rm ph} \ket{\phi_b=\beta}_{\rm ph}$}.
In each mode, the state evolves from $\ket{\phi=0}_{\rm ph}$ first into a squeezed state \cite{Ueda:1993} and then through macroscopic superpositions 
\cite{yurke1986, stoler1971}, eventually back to the phase state $\ket{\phi=\pi}_{\rm ph}$.

The prefactor $e^{it\chi_{ab}\hat{S}_z^a\hat{S}_z^b}$ in \eqref{eqn:psiT} entangles modes $a$ and $b$, 
with the following mechanism.
If $t$ is a rational fraction $m/q$ of $2t_{\pi}=2\pi/\chi_{ab}$, then the operator $e^{it\chi_{ab}\hat{S}_z^a\hat{S}_z^b}$ 
can be written as a sum of $q^2$ terms performing rotations in $a$ and $b$ modes 
(see Appendix \ref{app:special_states}).
Then acting with this operator on the state $\ket{\Psi_a,\Psi_b}$ one obtains a superposition of $q^2$ states.
If $q<\sqrt{N}$ and hence $\chi_{ab} t > 1/\sqrt{N}$, one can show that the reduced density matrix is a mixture of $q$ states, leading to an entropy of entanglement $S=\log (q)$ (see Appendix \ref{app:special_states}).
In particular at time $t=t_{\pi}$ we obtain
\begin{equation}
\chi=\chi_{ab}: \ \ket{\Psi(t_{\pi})}=\frac{\ket{\pi \pi}_{\rm ph}+\ket{ 0 \pi}_{\rm ph}+\ket{\pi 0}_{\rm ph}-\ket{00}_{\rm ph}}{2}
\label{eq:EvenCat}
\end{equation}
for $N$ even and
\begin{equation}
\chi=\chi_{ab}: \ \ket{\Psi(t_{\pi})}= \frac{\ket{\pi \pi}_{\rm ph}+i \ket{ 0 \pi}_{\rm ph}+i \ket{\pi 0}_{\rm ph}+\ket{00}_{\rm ph}}{2}
 \label{eq:OddCat}
\end{equation}
for $N$ odd.
For $N$ even, after measuring the phase $\phi\in \{0,\pi\}$ in the left well $b$, the state of the system $a$ is then projected onto a cat state  $\ket{\pi}_{\rm ph}-e^{i\phi}\ket{0}_{\rm ph}$.

In Fig.~\ref{fig:entropy}b-c we consider two other cases: $\chi/\chi_{ab}=0$ where only the $a$-$b$ entangling interaction is kept and $\chi/\chi_{ab}=0.5$. The latter value of $\chi/\chi_{ab}$
could in principle be obtained in optical potentials using a Feshbach resonance e.g.\ for ${}^{87}$Rb in states $\ket{0}=\ket{F=1,m_F=1}$ and $\ket{1}=\ket{F=2,m_F=-1}$ \cite{Oberthaler:2010,Sengstock:2004}, {or in a  variation of our scheme on an atom chip} \footnote{The parameter value 
$\chi=0.5 \chi_{ab}$ is in fact natural within another scheme
in which the components that interact are $1a$ and $1b$ (instead of {$1b$ and $0a$}). This scheme leads to the Hamiltonian {$H_{\rm int}^{\rm nl}  = \chi \bb{\hat{S}_z^{a}}^2+ \chi \bb{\hat{S}_z^{b}}^2  + \chi_{ab} \hat{S}_z^{a} \hat{S}_z^{b}$} with {$\chi$} defined as in equations \eqref{eq:def_chi_epsilon_sigma} and \eqref{eq:def_chi} while {$\chi_{ab}=2g_{11}\int |\phi_{1,a}|^2 |\phi_{1,b}|^2$}. In this case {$\chi=0.5 \chi_{ab}$}
corresponds to equal coupling constants and perfect overlap between $\phi_{1,a}$ and $\phi_{1,b}$.  This scheme is indeed realizable with microwave potentials {on an atomchip \cite{Treutlein2006}.}}.
Remarkably, although the entropy of entanglement remains the same \footnote{Note that the discussion of Appendix \ref{app:special_states} is independent on the ratio $\chi/\chi_{ab}$.}, very different macroscopic superpositions are obtained for different ratios $\chi/\chi_{ab}$. For example for $t=\pi/\chi_{ab}$ and $N$ even, we get
\begin{eqnarray}
\label{eq:phasestatespi}
\chi=0: &&  \ket{\Psi(t_\pi)}=\frac{\ket{00}_{\rm ph}+\ket{0\pi}_{\rm ph}+\ket{\pi0}_{\rm ph}-\ket{\pi\pi}_{\rm ph}}{2} \notag \\
 \chi=\frac{1}{2}\chi_{ab}: && \ket{\Psi(t_\pi)}=\frac{\ket{00}_{\rm ph}-i\ket{\pi\pi}_{\rm ph}}{\sqrt{2}}. 
\label{eq:GHZ}
\end{eqnarray}
To visualize these states, in the insets of Fig.~\ref{fig:entropy}
we show the full density matrix in the phase state basis
\begin{equation}
\label{eq:densitymatrixphase}
\rho_{\phi_a, \phi_b}=\bra{\phi_a \phi_b}\rho\ket{\phi_a \phi_b}
\end{equation}
with the phases $\phi_a$ and $\phi_b$ ranging in 
$[\pi/2,5\pi/2]$.
{We point out that a $\pi/2$-pulse in each condensate transforms the state \eqref{eq:GHZ} in a $a$-$b$ entangled GHZ-like state $|N,N\rangle_{\rm F} -i |0,0\rangle_{\rm F}$ with the notation \eqref{eq:notation_Fock}.}
The entangled macroscopic superpositions discussed here, although very interesting from the quantum information point of view, would probably be extremely challenging to observe. Indeed any source of decoherence, and in particular particle losses, should be avoided during the interaction time (typically a fraction of a second) \cite{Sinatra:1998,Pawlowski:2013}.

In the following, we focus on short time evolution $\chi_{ab} t \ll 1/\sqrt{N}$, in a regime in which we expect two-mode spin squeezing and EPR correlations. In Fig.~\ref{fig:entropy} this is the region in which the entropy grows monotonically.

\section{Spin squeezing and EPR entanglement}
\label{sec:short_times}

\subsection{EPR entanglement criterion}
\begin{figure*}
\begin{minipage}[l]{0.32\linewidth}
(a) $\chi=0$ \\
\includegraphics[width=\textwidth]{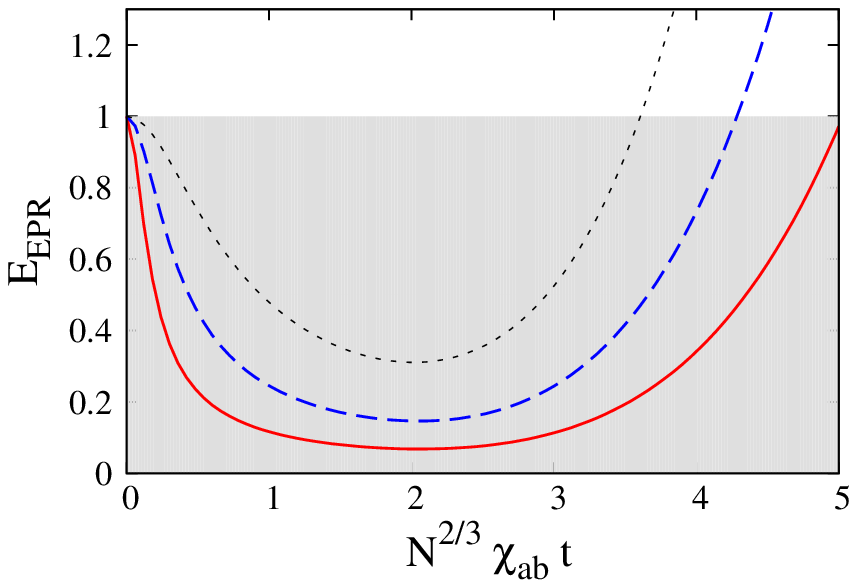} \\
\includegraphics[width=\textwidth]{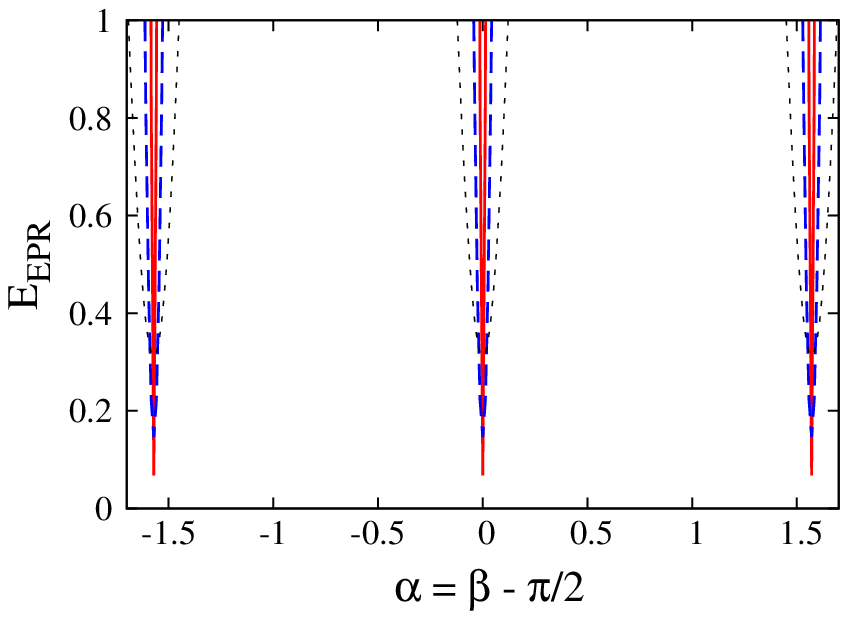}
\end{minipage}
\begin{minipage}[c]{0.32\linewidth}
(b) $\chi=\chi_{ab}$ \\
\includegraphics[width=\textwidth]{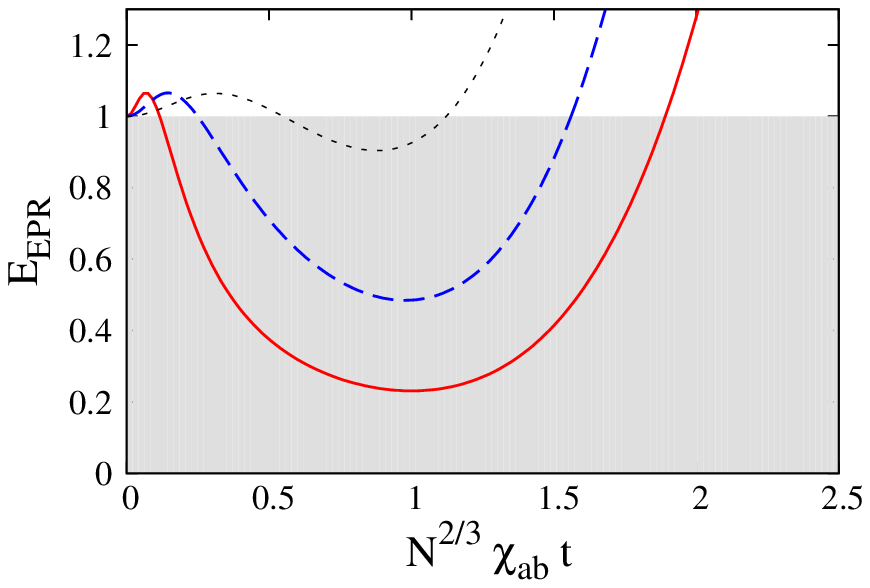}\\
\includegraphics[width=\textwidth]{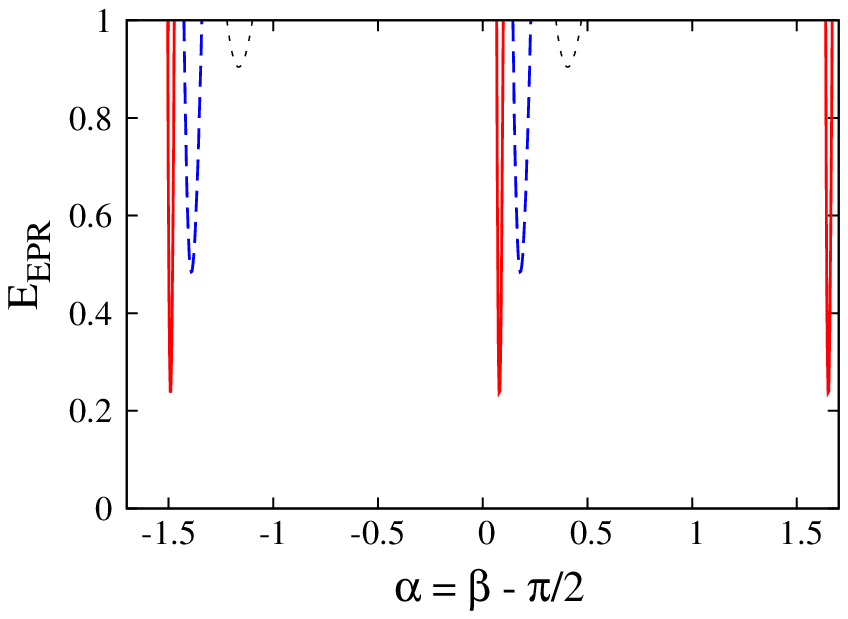}
\end{minipage}
\begin{minipage}[r]{0.32\linewidth}
(c) $\chi=0.5\chi_{ab}$ \\
\includegraphics[width=\textwidth]{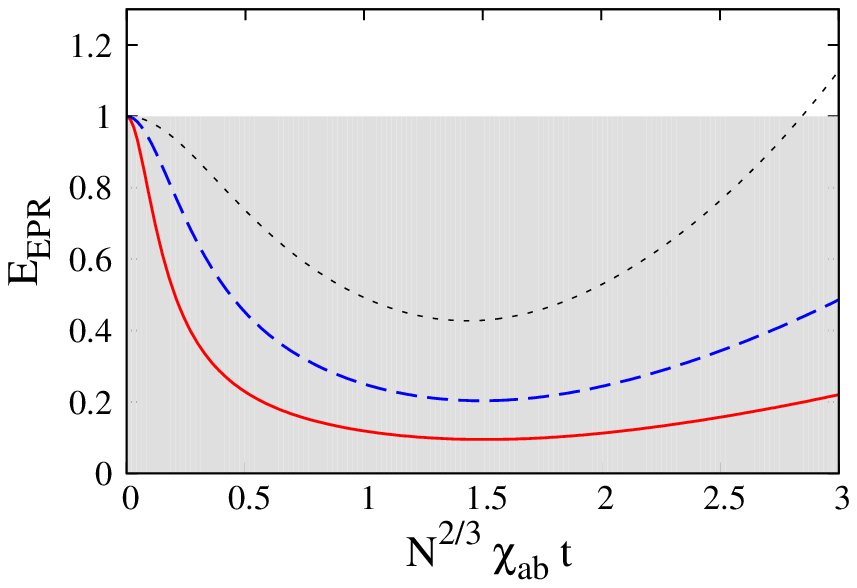}\\
\includegraphics[width=\textwidth]{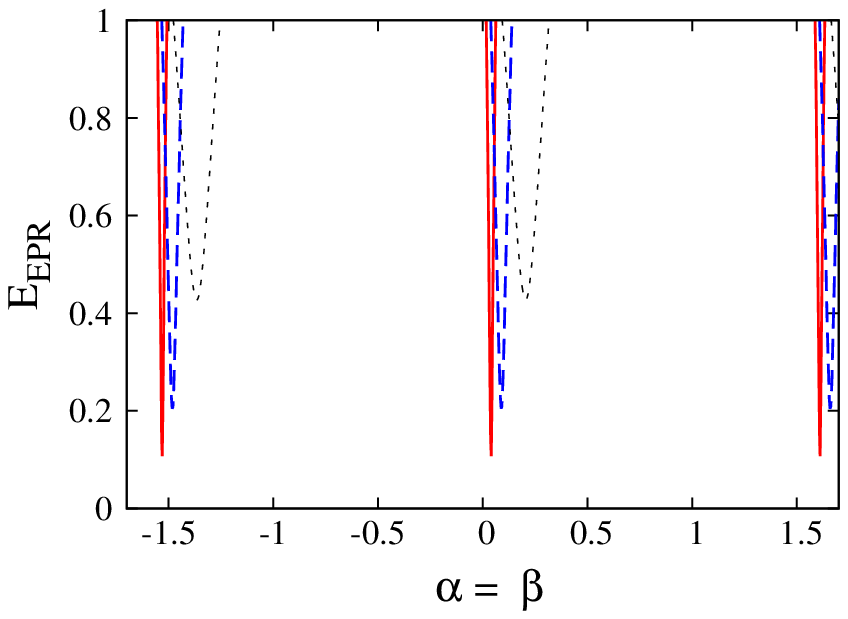}
\end{minipage}
\caption{(Color online) Top row: evolution of $E_{\rm EPR}$ \eqref{eq:E_EPR} for three different ratios $\chi/\chi_{ab}$. The number of atoms in each ensemble is $N=50$ (black dotted line), $N=500$ (blue dashed line) or $N=5000$ (red solid line). The quadrature angle $\alpha$ has been optimized numerically, while $\beta$ is fixed by the conditions \eqref{eq:cond_chi0},\eqref{eq:cond_chi1},\eqref{eq:cond_chi0.5}. Bottom row : dependence of $E_{\rm EPR}$ on the angle $\alpha$ while $\beta$ is still given by the conditions \eqref{eq:cond_chi0},\eqref{eq:cond_chi1},\eqref{eq:cond_chi0.5}. In these curves the time is fixed to the best time (minimum of $E_{\rm EPR}$) extracted from the top row curves although the time dependence is smooth in the interesting region \eqref{eq:squeezingregime}.
\label{fig:epr}} 
\end{figure*}
We imagine a situation where Alice (system $a$) and Bob (system $b$) can measure $\hat{X}_{a(b)}$ or $\hat{P}_{a(b)}$ on our bi-partite system, where $\hat{X}$ and $\hat{P}$ are two non commuting observables.
To quantify the entanglement of the state between Alice and Bob, we introduce the conditional variances \cite{walls_milburn_book_94,Wiseman2007,Cavalcanti2011,Opanchuk2012}:
\begin{eqnarray}
\Delta^2_{\rm inf}(\hat{X}_b) \equiv \meanv{\bb{\hat{X}_b-\hat{X}_b^{\rm inf}}^2}, \notag \\
\Delta^2_{\rm inf}(\hat{P}_b) \equiv \meanv{\bb{\hat{P}_b-\hat{P}_b^{\rm inf}}^2},
\end{eqnarray}
where $\hat{X}_b^{\rm inf}$ is an operator making an affine estimation of Bob's result for a measurement of $\hat{X}_b$ using Alice's measurement of $\hat{X}_a$
(and similarly for $\hat{P}_b^{\rm inf}$ that estimates Bob's result for $\hat{P}_b$ given Alice's result for $\hat{P}_a$), i.e.\
\be
\hat{X}_b^{\rm inf}=q_1 + q_2 \hat{X}_a, \qquad
\hat{P}_b^{\rm inf}=q_1' + q_2' \hat{P}_a.
\ee
Initially, the $a$ and $b$ systems are uncorrelated and
\be
\Delta^2(\hat{X}_b-\hat{X}_b^{\rm inf})=\Delta^2 \hat{X}_b+q_2^2 \, \Delta^2 \hat{X}_a
\ee
As the two sub-systems become more and more entangled, a measurement of $\hat{X}_a$ yields more and more information on $\hat{X}_b$ (in the extreme case of a perfect correlation one would get $\Delta^2(\hat{X}_b-\hat{X}_b^{\rm inf})=0$). Of course, the same picture can be drawn for $\Delta^2(\hat{P}_b-\hat{P}_b^{\rm inf})$. 
In a \textit{local hidden variable} theory the correlations between Alice and Bob would come from some hidden element of reality rather than from the non-locality of the quantum state. If we further ask that this theory is {\it locally compatible with quantum mechanics}, then Bob has his own quantum state that is not affected by Alice's measurements on her system. In this case $\hat{X}_b^{\rm inf}$ and $\hat{P}_b^{\rm inf}$ should be considered as numbers, as far as Bob is concerned, so that the Heisenberg uncertainty principle constrains the product $\Delta^2_{\rm inf}(\hat{X}_b) \Delta^2_{\rm inf}(\hat{P}_b)$.
If {this} constraint is violated,
\be
\Delta^2_{\rm inf}(\hat{X}_b) \Delta^2_{\rm inf}(\hat{P}_b) < \frac{1}{4} \left|\meanv{[\hat{X}_b,\hat{P}_b]}\right|^2,
\label{eq:EPRcriterion}
\ee
we have to admit that Alice's measurement changes Bob's state, which is {at the heart} of the EPR paradox.

To achieve \eqref{eq:EPRcriterion} in our situation, we use the best linear estimation of $\hat{X}_b$ using $\hat{X}_a$:
\be
\hat{X}_b^{\rm inf}=\meanv{\hat{X}_b}+\frac{\mbox{Covar}(\hat{X}_a,\hat{X}_b)}{\Delta^2 \hat{X}_a} \bb{\hat{X}_a-\meanv{\hat{X}_a}}
\ee
where $\mbox{Covar}(\hat{X}_a,\hat{X}_b)=\meanv{\hat{X}_a\hat{X}_b}-\meanv{\hat{X}_a}\meanv{\hat{X}_b}$. 

\subsection{EPR entanglement in our system}
For our system of two collective spins, we choose the non-commuting variables
\begin{eqnarray}
\hat{X}_a=\hat{S}_\alpha^a, \qquad \hat{P}_a=\hat{S}_{\alpha+\pi/2}^a, \notag \\
\hat{X}_b=\hat{S}_\beta^b, \qquad \hat{P}_b=\hat{S}_{\beta+\pi/2}^b.
\label{eq:genericquadratures}
\end{eqnarray}
Here $\hat{S}_\alpha^a$ and $\hat{S}_{\alpha+\pi/2}^a$ are spin operators obtained from $\hat{S}_y^a$ and $\hat{S}_z^a$ by linear combinations
\begin{eqnarray}
\hat{S}_\alpha^a &=&\cos \alpha\ \hat{S}_y^{a}   + \sin\alpha\ \hat{S}_z^{a} \notag \\
\hat{S}_{\alpha+\pi/2}^{a} &=&-\sin\alpha\ \hat{S}_y^{a}   + \cos\alpha\  \hat{S}_z^{a}
\label{eq:alpha}
\end{eqnarray}
with a commutator
\be
\left[\hat{S}_\alpha^a,\hat{S}_{\alpha+\pi/2}^a\right]=i\hat{S}_x^a
\ee
and similarly for $b$.

Rewriting \eqref{eq:EPRcriterion} using these operators yields the criterion:
\begin{widetext}
\begin{equation}
E_{\rm EPR}^2 \equiv \frac{ 4 \bb{\Delta^2 \hat{S}_\alpha^a \Delta^2 \hat{S}_\beta^b-\text{Covar}^2(\hat{S}_\alpha^a,\hat{S}_\beta^b)}\bb{\Delta^2 \hat{S}_{\alpha+\pi/2}^a \Delta^2 \hat{S}_{\beta+\pi/2}^b-\text{Covar}^2(\hat{S}_{\alpha+\pi/2}^a,\hat{S}_{\beta+\pi/2}^b)}}{\bb{\Delta^2\hat{S}_\alpha^a\Delta^2\hat{S}_{\alpha+\pi/2}^a} \left|\meanv{\hat{S}_x^b}\right|^2 }<1.
\label{eq:E_EPR}
\end{equation}
\end{widetext}
The quantum averages involved in the EPR criterion \eqref{eq:E_EPR} can be calculated analytically and are given in Appendix~\ref{app:averages}. As $E_{\rm EPR}$ depends on the angles $\alpha$ and $\beta$, the achievement of the inequality  \eqref{eq:E_EPR} in general requires a careful choice of the quadratures.

In Fig.~\ref{fig:epr}, top row, we show the evolution of $E_{\rm EPR}$ for optimized quadratures angles.
{In order to allow the comparison of curves with different atom number, the time has been rescaled 
using the scaling of the best squeezing time in the case of spin squeezing in a single condensate \cite{Ueda:1993}.}
In the bottom row we present the angular dependence of $E_{\rm EPR}$ at the best time extracted from the top row curves.

\paragraph{Case $\chi=0$.}
The quadrature optimization gives in this case
\be
\alpha=0, \quad \beta=\pi/2, 
\label{eq:cond_chi0}
\ee
corresponding to $\hat{X}_a=\hat{S}_y^a$ and $\hat{X}_b=\hat{S}_z^b$.
The condition \eqref{eq:cond_chi0} can be simply explained by integrating the equation of motion of $\hat{S}_y^a$
for $\chi_{ab} t \ll 1/\sqrt{N}$ and $\chi t \ll 1/\sqrt{N}$ \footnote{\label{foot:Sx} For $\chi_{ab} t \ll 1/\sqrt{N}$  and $\chi t \ll 1/\sqrt{N}$ the coherence in each ensemble between internal state $0$ and $1$ remains high so that $\meanv{\hat{S}_x^{a(b)}}\simeq N/2$.}:
\be
\hat{S}_y^a (t) = \hat{S}_y^a (0) + N t \bb{\chi \hat{S}_z^a - \frac{\chi_{ab}}{2} \hat{S}_z^b}
\label{eq:sy_t}
\ee
Indeed for $\chi=0$ and 
\begin{eqnarray}
1/N \ll \chi_{ab} t &\ll& 1/\sqrt{N} \notag \\
\chi t &\ll& 1/\sqrt{N} 
\label{eq:squeezingregime}
\end{eqnarray}
 the component $\hat{S}_y^a$ becomes an ``enlarged copy'' of 
$\hat{S}_z^b$ \cite{Frontiers:2012}. 

As shown in Fig.~\ref{fig:epr}a,  for times \eqref{eq:squeezingregime}  the system beats the classical limit and satisfies \eqref{eq:E_EPR}. 
{ Using  parameters as in the experiment of \cite{Treutlein:2010} ($N=10^3$ and $\chi_{ab}=0.5~\mathrm{s}^{-1}$), we find that maximal EPR entanglement is created for $t\simeq 20$~ms, well within reach of such experiments.
We note  that the angular width over which \eqref{eq:E_EPR} is satisfied decreases with the atom number, $6^\circ$ for $N=500$ and $1^\circ$ for $N=5000$, which is however still compatible with the experimental control achieved in \cite{Treutlein:2010,Ockeloen:2013}.}

\paragraph{Case $\chi = \chi_{ab}$.}
For $\chi \neq 0$ the squeezing Hamiltonian correlates $\hat{S}_y^b$ with $\hat{S}_z^b$ and hence
with $\hat{S}_y^a$. There is then {\it a priori} a competition between  the correlation $\hat{S}_y^a$-$\hat{S}_y^b$
and the correlation $\hat{S}_y^a$-$\hat{S}_z^b$.
The numerical optimization gives
\be
\alpha=\alpha_{opt}(t), \quad \beta=\alpha+\pi/2,
\label{eq:cond_chi1}
\ee
where $\alpha_{\rm opt}$ is a small angle approaching zero in the interesting time domain \eqref{eq:squeezingregime} when $N$ tends to infinity.
The results for this case are in Fig.~\ref{fig:epr}b.
As in the case $\chi=0$, the angular width over which \eqref{eq:E_EPR} is satisfied decreases with $N$.

\paragraph{Case $\chi = 0.5 \chi_{ab}$.}
In this case the quadrature optimization gives
\be
\alpha=\beta=\alpha_{opt}(t),
\label{eq:cond_chi0.5}
\ee
where $\alpha_{\rm opt}$ is a small angle approaching zero in the domain \eqref{eq:squeezingregime} when $N$ tends to infinity (see Fig.~\ref{fig:epr}c).
From equation \eqref{eq:sy_t} we have in this case:
\bea
\hat{S}_y^a(t)-\hat{S}_y^a(0)&=&-\bb{\hat{S}_y^b(t)-\hat{S}_y^b(0)}
\eea
As far as one can neglect the initial conditions, perfectly correlated quadratures are obtained choosing $\hat{X}_a=\hat{S}_y^a$ and  $\hat{X}_b=\hat{S}_y^b$ (and thus $\hat{X}_b^{\rm inf}=-\hat{S}_y^a$), corresponding to $\alpha=\beta=0$.

\begin{figure}[t]
\begin{minipage}[l]{\linewidth}
$\chi=0.5\chi_{ab}$ \\
\includegraphics[width=0.8\textwidth]{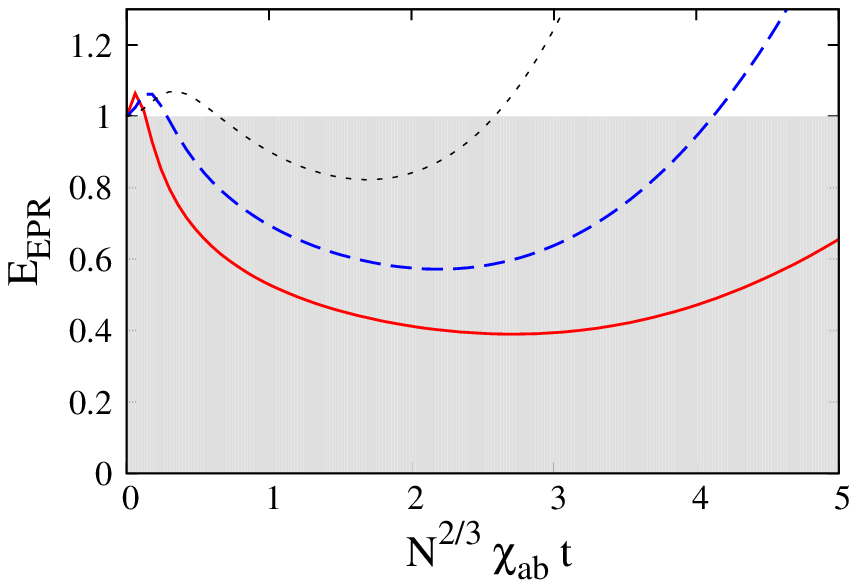} \\
\includegraphics[width=0.8\textwidth]{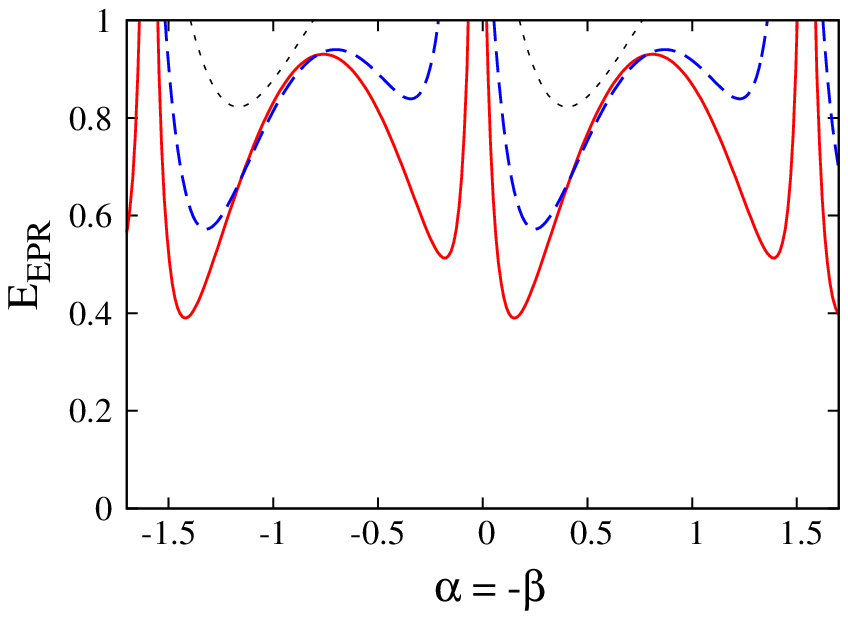}
\end{minipage}
\caption{(Color online) Top: evolution of $E_{\rm EPR}$ for $\chi=0.5\chi_{ab}$ and quadratures chosen according to \eqref{eq:cond_chi0.5_nopt}. The atom numbers are the same as in Fig.~{\ref{fig:epr}}. Bottom: at the best time,  dependence of $E_{\rm EPR}$ on the angle $\alpha$ while $\beta$ is still given by the conditions {\eqref{eq:cond_chi0.5_nopt}.}
\label{fig:eprMagic}} 
\end{figure}

The case $\chi=0.5\chi_{ab}$ is ``special'' in the sense that, if we expand $E_{\rm EPR}$ for times 
\eqref{eq:squeezingregime}, the leading order in $(Nt)$ (order four) is identically zero independently of the quadrature angles. This explains the fact that both $E_{\rm EPR}$ and the entanglement witness $E_{\rm ent}$ defined in (\ref{eq:entanglement}) {below} take very small values in this case although $\chi \neq 0$ (see Fig.~\ref{fig:epr} and Fig.~\ref{fig:entanglement}).
Over the three cases we have considered, $\chi = 0.5 \chi_{ab}$ is also the only one in which we find a solution,
that, although different from the global minimum of $E_{\rm EPR}$ over the quadrature angle, has
a smooth angular dependence of $E_{\rm EPR}$. This is obtained choosing
\be
\alpha=\alpha_{opt}, \quad  \beta=-\alpha,
\label{eq:cond_chi0.5_nopt}
\ee
and the result is shown in Fig.~\ref{fig:eprMagic}.

\subsection{Entanglement criterion}
We defined the EPR-entanglement criterion in \eqref{eq:EPRcriterion}.
Note that an ordinary {\it entanglement} criterion between ensembles $a$ and $b$ would be
\be
E_{\rm ent}^2 \equiv \frac{\Delta^2_{\rm inf}(\hat{X}_b) \Delta^2_{\rm inf}(\hat{P}_b) }{\Delta^2(\hat{X}_b) \Delta^2(\hat{P}_b)}< 1.
\label{eq:entanglement}
\ee
Indeed for two uncorrelated systems $\text{Cov}\bb{\hat{X}_a,\hat{X}_b}=0$, $\Delta^2_{\rm inf}(\hat{X}_b)=\Delta^2(\hat{X}_b)$  and  \eqref{eq:entanglement} becomes an equality. We note that 
while the EPR-entanglement criterion \eqref{eq:EPRcriterion} implies the entanglement criterion \eqref{eq:entanglement}, the converse is not true.

\begin{figure*}
\begin{minipage}[l]{0.32\linewidth}
(a) $\chi=0$ \\
\includegraphics[width=\textwidth]{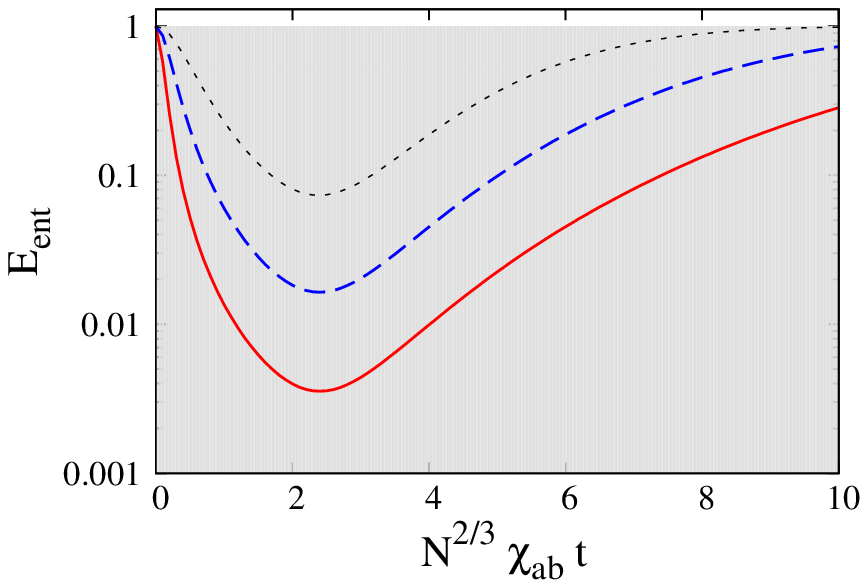} \\
\includegraphics[width=\textwidth]{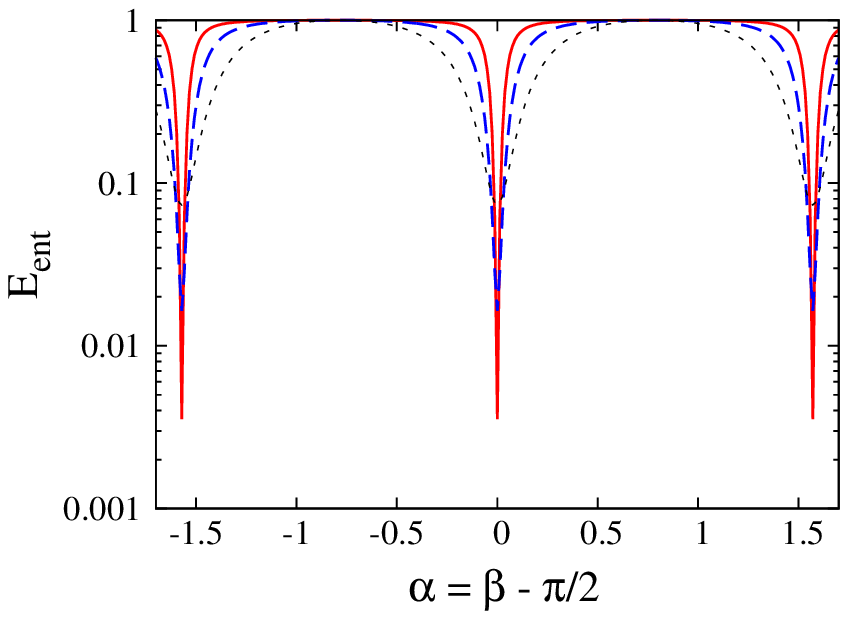}
\end{minipage}
\begin{minipage}[c]{0.32\linewidth}
(b) $\chi=\chi_{ab}$ \\
\includegraphics[width=\textwidth]{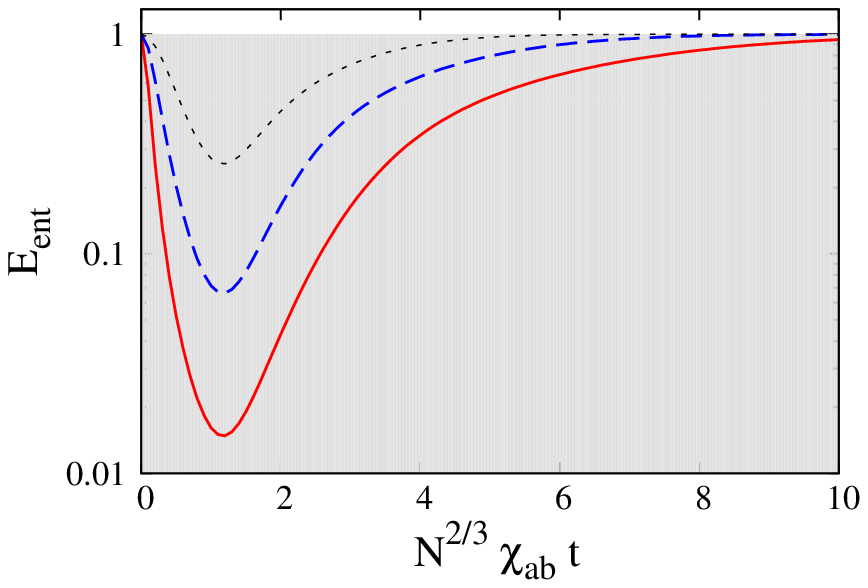}\\
\includegraphics[width=\textwidth]{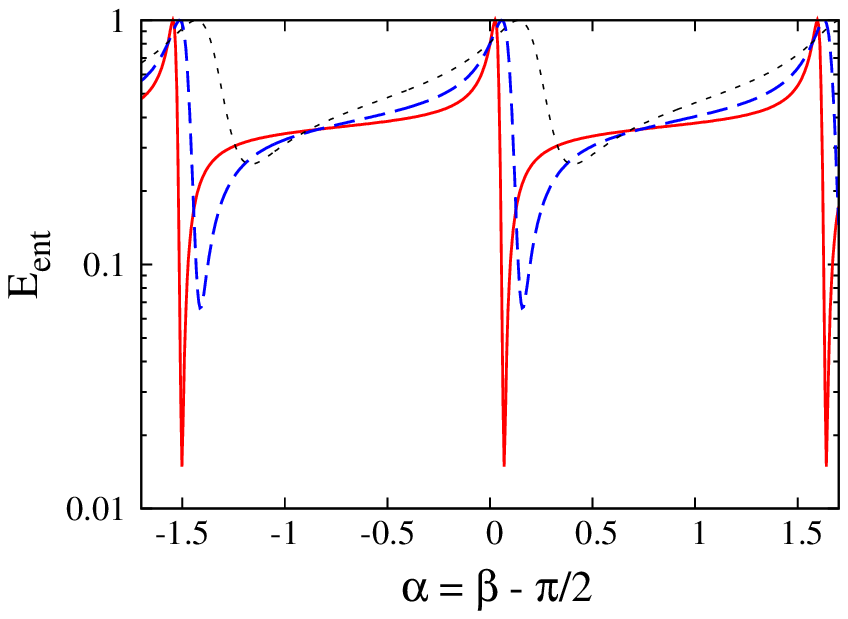}
\end{minipage}
\begin{minipage}[r]{0.32\linewidth}
(c) $\chi=0.5\chi_{ab}$ \\
\includegraphics[width=\textwidth]{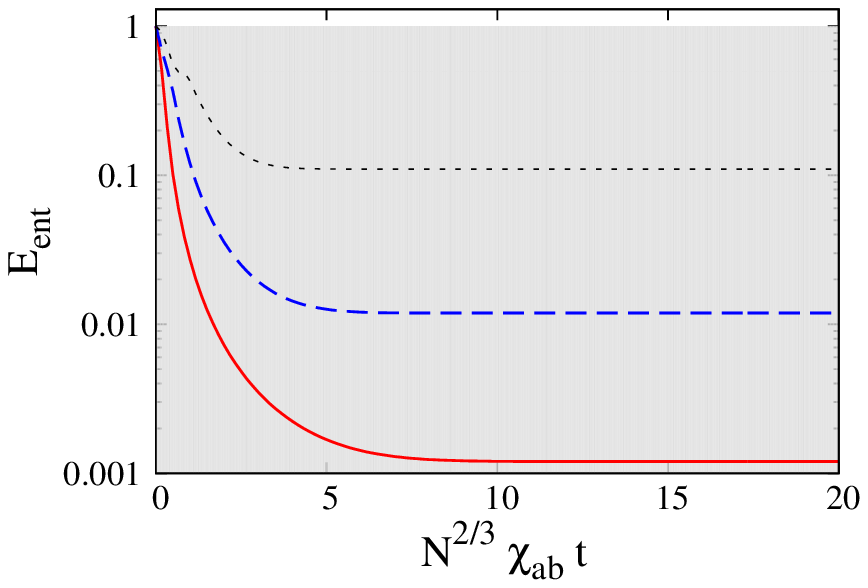}\\
\includegraphics[width=\textwidth]{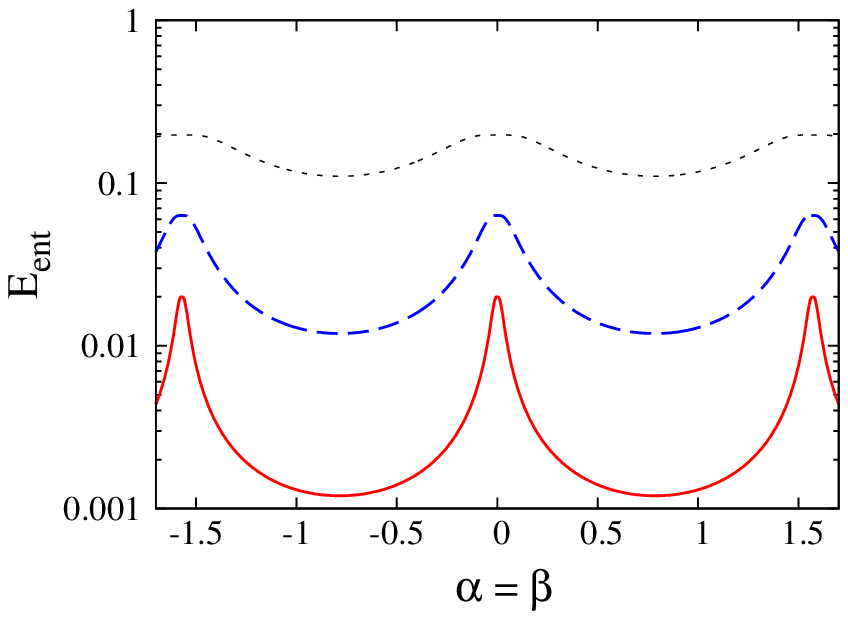}
\end{minipage}
\caption{\label{fig:entanglement}
(Color online) Top row: evolution of the entanglement witness $E_{\rm ent}$ \eqref{eq:entanglement} for three different ratios $\chi/\chi_{ab}$. The number of atoms in each ensemble is $N=50$ (black dotted line), $N=500$ (blue dashed line), and $N=5000$ (red solid line). The quadrature angle $\alpha$ has been optimized numerically, while $\beta$ is fixed by the conditions \eqref{eq:cond_chi0},\eqref{eq:cond_chi1},\eqref{eq:cond_chi0.5}. Bottom row: dependence of $E_{\rm ent}$ on the angle $\alpha$ while $\beta$ is still given by the conditions \eqref{eq:cond_chi0},\eqref{eq:cond_chi1},\eqref{eq:cond_chi0.5}. In these curves the time is fixed to the best time (minimum of $E_{\rm ent}$) extracted from the top row curves.}
\end{figure*}

We find that very strong $a$-$b$ correlations build up in our system so that the inferred variances become very small compared to the original variances. This holds for a wide choice of quadrature angles. We show an example in Fig.~\ref{fig:entanglement} for three values of the ratio $\chi/\chi_{ab}$.

\subsection{Squeezing versus EPR entanglement}
Finally one may ask whether the squeezing of each BEC resulting from the $\chi \bb{\hat{S}^a_z}^2$ 
and $\chi \bb{\hat{S}^b_z}^2$ terms in the Hamiltonian \eqref{eq:HamiltonianSimple} is affected by the non-local entanglement between condensates $a$ and $b$. 

In general if one system is entangled with another system that is not measured, the purity of its state is degraded
and so its quantum correlations. Based on this argument 
we thus expect the squeezing of the quadrature in $a$ to be deteriorated by the EPR correlations, which
is what we investigate in this subsection.

\begin{figure}[t]
\begin{minipage}[l]{\linewidth}
(a) $\chi=0.5\chi_{ab}$ \\
\includegraphics[width=0.8\textwidth]{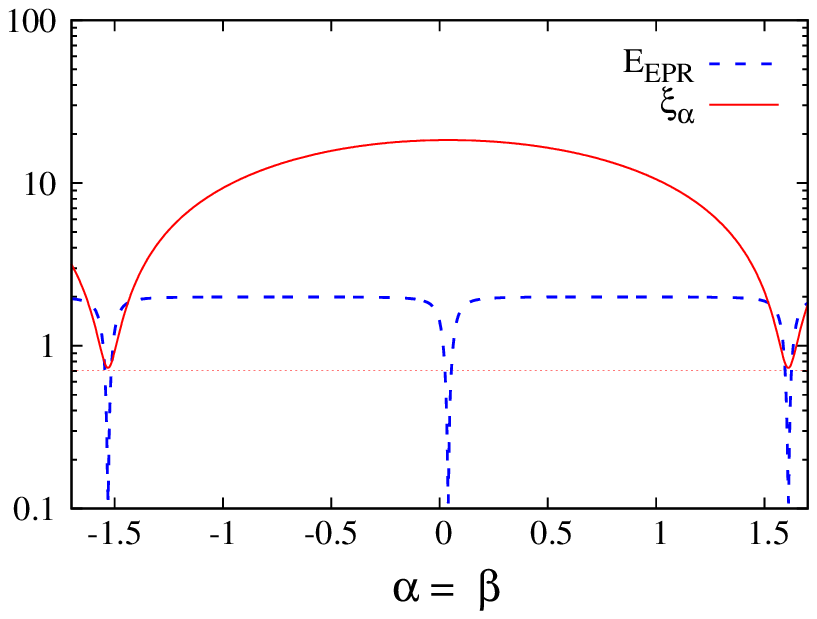} \\
(b) $\chi=\chi_{ab}$ \\
\includegraphics[width=0.8\textwidth]{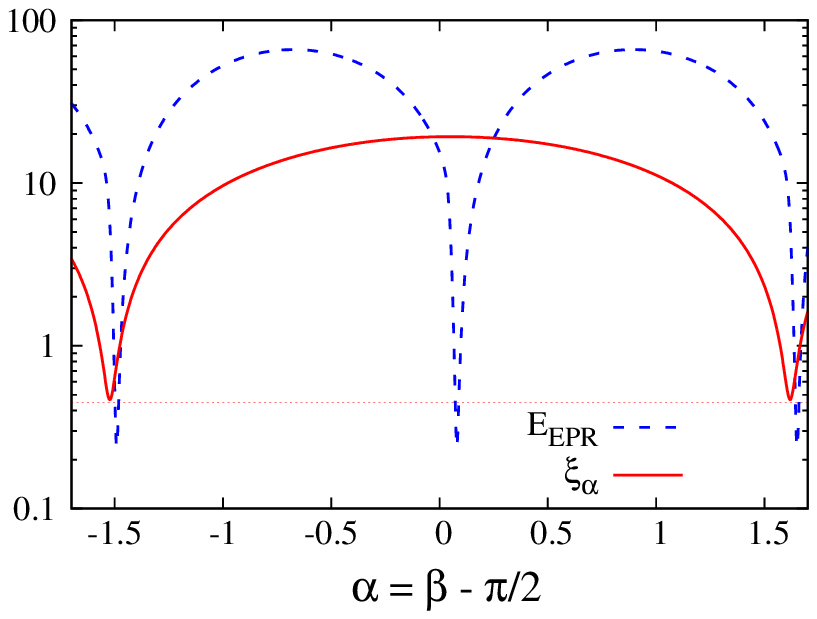}
\end{minipage}
\caption{(Color online) Angular dependence of the squeezing parameter $\xi_{\alpha}$ \eqref{eq:xi} (solid red line)
together with $E_{\rm{EPR}}$ (dashed blue line, already shown in Fig.\ref{fig:epr} bottom row), for two values
of the ratio $\eta=\chi_{ab}/2\chi$. 
For both $\xi_\alpha$ and $E_{\rm EPR}$ the time is fixed to the time minimizing $E_{\rm EPR}$ like in Fig. \ref{fig:epr}.
The atom number is $N=5000$ in each condensate. 
The horizontal red dotted line shows the analytical value of the squeezing limit \eqref{eq:xiBEST}.
\label{fig:xi_epr}}
\end{figure}

For the quadrature $\hat{S}_\alpha^a$ introduced in equation \eqref{eq:alpha} we define the squeezing parameter 
$\xi_{\alpha}$ \cite{Wineland:1994}:
\be
\xi_{\alpha} = \frac{\sqrt{N \Delta^2 \hat{S}_\alpha^a }}{\left|\meanv{\hat{\vec{S}}^a}\right|}.
\label{eq:xi}
\ee
The system is squeezed if $\xi_{\alpha}<1$. 
In the time regime given by \eqref{eq:squeezingregime}, one has
$\left|\meanv{\hat{\vec{S}}^a}\right|\simeq N/2$, so that we can focus on $\Delta^2  \hat{S}_\alpha^a$. 
From the definition of $\hat{S}_\alpha^a$ \eqref{eq:alpha} and equation \eqref{eq:sy_t} giving $\hat{S}_y^a (t)$,
one can compute the values of the quadrature angle for which $\Delta^2  \hat{S}_\alpha^a$ reaches its extrema
 \be
 \tan\, 2\alpha = \frac{\meanv{\hat{S}_z^a\hat{S}_y^a+\hat{S}_y^a\hat{S}_z^a} }
 { \meanv{ \bb{\hat{S}_y^a}^2} -\meanv{\bb{\hat{S}_z^a}^2}} = \frac{4 \eta}{\chi_{ab} N t \bb{1 + \eta^2}}
 \label{eq:tan2_alpha}
 \ee
 where 
 \be
 \eta = \chi_{ab}/2\chi. 
 \ee
Note that from \eqref{eq:squeezingregime}, $ \tan\, 2\alpha \ll 1$. Linearizing equation \eqref{eq:tan2_alpha} we find that the minimum of $\Delta^2  \hat{S}_\alpha^a$
is obtained for
\be
\alpha_{\rm{min}} = -\pi/2 + \frac{2 \eta}{N \chi_{ab} t \bb{1 + \eta^2}}
\ee
Expanding $\Delta^2  \hat{S}_\alpha^a$ in powers of $1/N \chi_{ab} t $ we finally obtain
\be
\xi_{\rm{min}} =\frac{\eta}{\sqrt{1+\eta^2}}  + O\bb{\frac{1}{\bb{N \chi_{ab} t }^2} }
\label{eq:xiBEST}
\ee
 This equation shows that the non-local entanglement ($\eta \neq 0$) introduces a squeezing limit
 that is independent on the particle number. The limit \eqref{eq:xiBEST} is shown as a dotted horizontal line
 in Fig.\ref{fig:xi_epr}.
Physically, the limitation of the squeezing comes from the fact that when $\chi_{ab}\neq0$ ($\eta \neq 0$), the state of 
$a$ after tracing out the system $b$ is not anymore a unique squeezed state but rather a mixture of squeezed states rotated by an angle depending on the state of $b$ \footnote{This can be seen by (i) writing the initial state as 
{$|\Psi(0)\rangle = \frac{1}{2^{N/2}} \sum_{n_b}\sqrt{\binom{N}{n_b}} \ket{n_b}_F\ket{\phi_a=0}_{\rm ph} $}, where we expanded only the phase states of $b$ in the Fock basis, (ii) evolving this state with the Hamiltonian 
\eqref{eq:HamiltonianSimple} and (iii) tracing out the system $b$.}.
From this figure we remark that although squeezing in each condensate is limited,  it can however coexist with EPR entanglement for experimentally relevant parameters.

\section{Conclusion}
We consider a scheme that allows to entangle two {spatially} separated atomic ensembles using {collisional} interactions. The system we propose is a pair of bimodal condensates in state dependent traps as could be realized in an optical trap or using microwave traps on an atom chip. Within a four mode approximation, we find a very rich physics. At long evolution times, the system evolves into macroscopic superpositions entangling the two ensembles $a$ and $b$. We have shown that these states take a simple form if expressed in the phase states basis. Correspondingly, at some remarkable times, the entanglement entropy takes simple values. At short times the system exhibits EPR-like entanglement that could be revealed by variance and covariance measurements on the collective spin components. We compared criteria for EPR-entanglement and simple entanglement. We show that they are both satisfied for a wide range of experimental parameters in our system, the EPR-entanglement being however more demanding than simple entanglement because of its sensitivity to the choice of quadratures. Finally we proved that the best achievable squeezing in each BEC is limited to a finite value by the non-local entanglement between $a$ and $b$.

We expect that the macroscopic superpositions that we predict at long times will be very sensitive to decoherence while the non-local entanglement at short times might be more accessible as it is the case for spin-squeezing in a single ensemble \cite{LiYun:2008}. Further studies including in particular the spatial dynamics of the condensates wave functions \cite{YunLi:2009} will be done in order to confirm the conclusion of the present work and to allow a detailed comparison with future experiments. During the preparation of the present manuscript we became aware of a preprint discussing ideas closely related to section IV of our paper \cite{Byrnes:2013}.

\begin{acknowledgments}
\label{sec:acknowl}
We acknowledge support from IFRAF-CNano \^Ile de France, {from the european project QIBEC, and from the Swiss National Science Foundation.}
\end{acknowledgments}

\appendix

\section{Why $\log (q)$?}
\label{app:special_states}
Here we show how to decompose the operator $\exp \bb{i2 \pi m/q\, \hat{S}_z^a\hat{S}_z^b}$ into a sum of simple rotations in systems $a$ and $b$ (see equation \eqref{eqn:decomposition}). We will use this result to calculate the global state and the entropy of entanglement at special times $\chi_{ab}t= 2\pi m/q$.

We consider $N$ even \footnote{The proof for $N$ odd is obtained by mapping {$\hat{S}_z^a\mapsto\hat{S}_z^a+1/2$.}}
and we introduce the notation 
\be
x\equiv\exp \bb{i2\pi m/q}.
\ee
First, we will prove the identity
\be
\sum_{p, r=0}^{q-1}x^{- \bb{ \hat{S}_z^a-p}\bb{\hat{S}_z^b-r}} = q.
\label{eqn:identity1}
\ee
It is {sufficient} to  check that \eqref{eqn:identity1} holds  in the eigenbasis of operators $\hat{S}_z^a$ and $\hat{S}_z^b$.
For $N$ even, both $\hat{S}_z^a$ and $\hat{S}_z^b$ have integer eigenvalues, which we denote $\delta n_a$ and $\delta n_b$, respectively.
For each pair of eigenvalues $\delta n_a$, $\delta n_b$ we obtain
\begin{equation}
\sum_{p, r=0}^{q-1}x^{- \bb{\delta n_a-p}\bb{\delta n_b-r}} = \sum_{p^{\prime}, r^{\prime}=0}^{q-1}x^{- p^{\prime}r^{\prime}} 
\label{eqn:dec_onBasis}
\end{equation}
where we redefined the summation indices $p^{\prime}\equiv (p-\delta n_a)$ mod $q$ 
and {$r^\prime\equiv (r-\delta n_b)$ mod $q$}. 
However, from the properties of  the roots (of degree $q$) of unity, we know that  $\sum_{r^{\prime}=0}^{q-1}x^{-p^{\prime}r^{\prime}} = q\,\delta_{p^{\prime}, 0}$, where $\delta_{i, j}$ is the Kronecker delta. Hence, we have $ \sum_{p^{\prime}, r^{\prime}=0}^{q-1}x^{- p^{\prime}r^{\prime}} =q$, which proves the equation \eqref{eqn:identity1}.

To decompose $\exp \bb{i 2\pi m/q\hat{S}_z^a\hat{S}_z^b}$ we use additionally the relation
\begin{equation}
x^{-pr}x^{p \hat{S}_z^b+r \hat{S}_z^a}=x^{\hat{S}_z^a\hat{S}_z^b} x^{- \bb{ \hat{S}_z^a-p}\bb{\hat{S}_z^b-r}} .
\label{eqn:keyId}
\end{equation}
After summing both sides of Eq.~\eqref{eqn:keyId} over $p, r=0, 1, 2, \ldots, q-1$ we get:
\begin{eqnarray}
\sum_{p, r=0}^{q-1} x^{-pr}x^{p \hat{S}_z^b+r \hat{S}_z^a} &=&x^{\hat{S}_z^a\hat{S}_z^b} \sum_{p, r=0}^{q-1}x^{- \bb{ \hat{S}_z^a-p}\bb{\hat{S}_z^b-r}} \notag \\
& \overset{ \text{(A.2)}}{=}& x^{\hat{S}_z^a\hat{S}_z^b} \;q.
\label{eq:helpKeyId}
\end{eqnarray}
Finally, dividing Eq. \eqref{eq:helpKeyId} by $q$ we obtain the desired  decomposition 
\begin{equation}
e^{i\frac{2\pi m}{q}\hat{S}_z^a\hat{S}_z^b} =\frac{1}{q}\sum_{p, r=0}^{q-1}x^{-pr}\,e^{i\frac{2\pi m}{q} \, p \, \hat{S}_z^b}\,e^{i\frac{2\pi m}{q} \, r  \hat{S}_z^a}.
\label{eqn:decomposition}
\end{equation}
The entangling rotation in $a$ and $b$ is thus rewritten as a sum of simultaneous rotations by all multiples of an angle $2\pi /q$ in the $a$ and $b$ systems with coefficients {$\frac{1}{q} x^{-pr}=\frac{1}{q}\exp \bb{-i 2\pi m\,  p \,r\, /q}$}.
Acting with this operator on a product state $\ket{\Psi_a,\Psi_b}$ we get
\begin{eqnarray*}
e^{i\frac{2\pi m}{q}\hat{S}_z^a\hat{S}_z^b}\ket{\Psi_a,\Psi_b}&=& \frac{1}{q}\sum_{p, r=0}^{q-1} x^{-pr}e^{i\frac{2\pi m}{q}\bb{ p \hat{S}_z^a+ r \hat{S}_z^b} }\ket{\Psi_a,\Psi_b}\\
 &=& \frac{1}{\sqrt{q}}\sum_{p=0}^{q-1} e^{i\frac{2\pi m}{q} p \hat{S}_z^a}\ket{\Psi_a,\Psi_b^{(p)}},
\end{eqnarray*}
where $\ket{\Psi_b^{(p)}} = \frac{1}{\sqrt{q}}\sum_{r=0}^{q-1}  x^{-pr} e^{i\frac{2\pi m}{q} r \hat{S}_z^b}\ket{\Psi_b}$.

From now on we focus on "long" evolution times, where the angle $1/q > 1/\sqrt{N}$.
Under this condition any phase state $\ket{\phi}_{\rm ph}$ is almost orthogonal to its rotated version {$\exp\bb{i\frac{2\pi}{q} r \hat{S}_z^b}\ket{\phi}_{\rm ph}$.}
This is also the case {for} the evolved state $\ket{\Psi_b}=e^{-i \chi t \bb{\hat{S}_z^b}^2}\ket{0}_{\rm ph}$ considered in the main text:
\begin{eqnarray}
\bra{\Psi_b}e^{- i\frac{2\pi}{q} r\hat{S}_z^b} \ket{\Psi_b}&=& _{\rm ph}\bra{0}e^{i \chi t \bb{\hat{S}_z^b}^2} e^{- i\frac{2\pi}{q} r\hat{S}_z^b} e^{-i \chi t \bb{\hat{S}_z^b}^2}\ket{0}_{\rm ph} \notag \\
 &= & _{\rm ph}\bra{0} e^{- i\frac{2\pi}{q} r\hat{S}_z^b} \ket{0}_{\rm ph} \simeq \delta_{r, 0}.
 \label{eqn:PsiB}
\end{eqnarray}
From \eqref{eqn:PsiB} we see that states  $\ket{\Psi_b^{(p)}} $ form an almost orthonormal basis:
\begin{eqnarray}
\bra{\Psi_b^{(p^{\prime})}} \Psi_b^{(p)} \rangle &=& \frac{1}{q}\bra{\Psi_b}\sum_{r, r^{\prime}} x^{p^{\prime}r^{\prime}-pr}e^{i\frac{2\pi m}{q} \bb{r-r^{\prime}} \hat{S}_z^b}\ket{\Psi_b} \notag \\
 &=&\frac{1}{q} {\sum_{r}} x^{-r \bb{p-p^{\prime}}}  = \delta_{p, p^{\prime}}.
\end{eqnarray}
With this we can  compute the reduced density matrix for system $a$:
\begin{eqnarray}
\rho_a &=&  {\rm Tr}_b \rho = \sum_{p} \bra{\Psi_b^{(p)}} \rho \ket{\Psi_b^{(p)}} \notag \\
 & = &\frac{1}{q} \sum_{p=0}^{q-1} e^{  i\frac{2\pi}{q} p\hat{S}_z^a} \ \ket{\Psi_a}\bra{\Psi_a}e^{-i\frac{2\pi}{q} p \hat{S}_z^a}.
\end{eqnarray}
{This mixed state} indeed leads us to the value of the entropy close to $\log (q)$ due to the orthogonality of states $e^{  i\frac{2\pi}{q} p\hat{S}_z^a} \ket{\Psi_a}$ for different $p$.

An important example is the case $q=2$, which corresponds to $t_{\pi} = \frac{\pi}{\chi_{ab}}$.
According to \eqref{eqn:decomposition} we can write
\begin{equation}
e^{i\pi \hat{S}_z^a\hat{S}_z^b} =\frac{1}{2}\bb{1+e^{i\pi \hat{S}_z^a}+e^{i\pi \hat{S}_z^b}-e^{i\pi \bb{\hat{S}_z^b+\hat{S}_z^a}}}.
\label{eqn:catCreation}
\end{equation}
In the case $\chi=\chi_{ab}$ the state $\ket{\Psi_a, \Psi_b}$ is equal to $\ket{\pi\pi}_{\rm ph}$.
Thus according to \eqref{eqn:catCreation} and  \eqref{eqn:psiT} the global state of the system is
\begin{eqnarray*}
\ket{\Psi (t_{\pi})}&=&e^{i\pi \hat{S}_z^a\hat{S}_z^b}\ket{\pi\pi}_{\rm ph}\\
 &=&\frac{1}{2}\bb{\ket{\pi\pi}_{\rm ph}+\ket{0\pi}_{\rm ph}+\ket{\pi0}_{\rm ph}-\ket{00}_{\rm ph}}.
 \end{eqnarray*}
In such a way one can reconstruct all the states shown in insets of the Fig.~\ref{fig:entropy}, corresponding to different $q$ and ratios $\chi/\chi_{ab}$.

\section{Quantum averages}
\label{app:averages}

The quantum averages used to compute the EPR criterion are listed below:

\begin{eqnarray}
 \meanv{\hat{S}_x^a}&=& \frac{N}{2} \cos^{N-1} \bb{ \chi t} \,\cos^{N} \bb{ \chi_{ab} t/2} \label{eqn:sX}, \notag \\ 
 \meanv{\hat{S}_y^a}&=& 0,\quad \quad \notag
 \meanv{\hat{S}_z^a}= 0.
 \end{eqnarray}
 
 \begin{eqnarray}
 \meanv{ \bb{\hat{S}_y^a} ^2 }&=&\frac{N}{4} \notag\\ &+&\frac{N(N-1)}{8}\bb{1- \cos^{N} \bb{\chi_{ab}t}\,\cos^{N-2} \bb{2\chi t}},  \notag \\
 \meanv{ \bb{\hat{S}_z^a} ^2 }&=&\frac{N}{4}, \notag \\
 \meanv{ \hat{S}_z^a \hat{S}_z^b }&=& 0, \label{eqn:SzAB} \notag \\
 \nonumber\meanv{ \hat{S}_y^a \hat{S}_y^b }&=&\frac{N^2}{8} \Big{(}\cos^{2N-2} \bb{\chi + \chi_{ab}/2) t} \notag \\ && \qquad \qquad - \cos^{2N-2}\bb{ (\chi - \chi_{ab}/2) t}\Big{)}, \notag
\label{eqn:SyAB}
 \end{eqnarray}
 
 \begin{eqnarray}
 \meanv{ \hat{S}_y^a \hat{S}_z^b }&=& -\frac{N^2}{4} \sin(\chi_{ab}t/2) \cos^{N-1}(\chi_{ab}t/2) \cos^{N-1}(\chi t),\notag \\
 \meanv{ \{\hat{S}_y^a,\hat{S}_z^a\} }&=& \frac{N (N-1)}{2} \sin(\chi t) \cos^{N}(\chi_{ab}t/2) \cos^{N-2}(\chi t). \notag
\end{eqnarray}

The variances and covariances of the generic quadratures \eqref{eq:genericquadratures} are 
\begin{eqnarray}
\Delta^2\hat{S}_\alpha^a&=&\cos^2\alpha\meanv{\bb{\hat{S}_y^a}^2}+\sin^2\alpha\meanv{\bb{\hat{S}_z^a}^2}\notag\\&&\qquad\qquad\qquad +\sin\alpha\cos\alpha\meanv{ \{\hat{S}_y^a,\hat{S}_z^a\} }, \notag \\
\Delta^2\hat{S}_\beta^b&=&\cos^2\beta\meanv{\bb{\hat{S}_y^b}^2}+\sin^2\beta\meanv{\bb{\hat{S}_z^b}^2}\notag\\&&\qquad\qquad\qquad +\sin\beta\cos\beta\meanv{ \{\hat{S}_y^b,\hat{S}_z^b\} }, \notag \\
\text{Covar}(\hat{S}_{\alpha}^a,\hat{S}_{\beta}^b)&=&\cos\alpha\cos\beta\meanv{\hat{S}_y^a \hat{S}_y^b}\notag\\&+&\sin\alpha\cos\beta\meanv{\hat{S}_y^b \hat{S}_z^a}+\sin\beta\cos\alpha\meanv{\hat{S}_y^a \hat{S}_z^b}. \notag
\end{eqnarray}



\begin{thebibliography}{36}
\expandafter\ifx\csname natexlab\endcsname\relax\def\natexlab#1{#1}\fi
\expandafter\ifx\csname bibnamefont\endcsname\relax
  \def\bibnamefont#1{#1}\fi
\expandafter\ifx\csname bibfnamefont\endcsname\relax
  \def\bibfnamefont#1{#1}\fi
\expandafter\ifx\csname citenamefont\endcsname\relax
  \def\citenamefont#1{#1}\fi
\expandafter\ifx\csname url\endcsname\relax
  \def\url#1{\texttt{#1}}\fi
\expandafter\ifx\csname urlprefix\endcsname\relax\def\urlprefix{URL }\fi
\providecommand{\bibinfo}[2]{#2}
\providecommand{\eprint}[2][]{\url{#2}}

\bibitem[{\citenamefont{Riedel et~al.}(2010)\citenamefont{Riedel, B{\"o}hi, Li,
  H{\"a}nsch, Sinatra, and Treutlein}}]{Treutlein:2010}
\bibinfo{author}{\bibfnamefont{M.}~\bibnamefont{Riedel}},
  \bibinfo{author}{\bibfnamefont{P.}~\bibnamefont{B{\"o}hi}},
  \bibinfo{author}{\bibfnamefont{Y.}~\bibnamefont{Li}},
  \bibinfo{author}{\bibfnamefont{T.}~\bibnamefont{H{\"a}nsch}},
  \bibinfo{author}{\bibfnamefont{A.}~\bibnamefont{Sinatra}}, \bibnamefont{and}
  \bibinfo{author}{\bibfnamefont{P.}~\bibnamefont{Treutlein}},
  \bibinfo{journal}{Nature} \textbf{\bibinfo{volume}{464}},
  \bibinfo{pages}{1170} (\bibinfo{year}{2010}).

\bibitem[{\citenamefont{Gross et~al.}(2010)\citenamefont{Gross, Zibold,
  Nicklas, Esteve, and Oberthaler}}]{Oberthaler:2010}
\bibinfo{author}{\bibfnamefont{C.}~\bibnamefont{Gross}},
  \bibinfo{author}{\bibfnamefont{T.}~\bibnamefont{Zibold}},
  \bibinfo{author}{\bibfnamefont{E.}~\bibnamefont{Nicklas}},
  \bibinfo{author}{\bibfnamefont{J.}~\bibnamefont{Esteve}}, \bibnamefont{and}
  \bibinfo{author}{\bibfnamefont{M.}~\bibnamefont{Oberthaler}},
  \bibinfo{journal}{Nature} \textbf{\bibinfo{volume}{464}},
  \bibinfo{pages}{1165} (\bibinfo{year}{2010}).

\bibitem[{\citenamefont{Hamley et~al.}(2012)\citenamefont{Hamley, Gerving,
  Hoang, Bookjans, and Chapman}}]{Hamley2012}
\bibinfo{author}{\bibfnamefont{C.~D.} \bibnamefont{Hamley}},
  \bibinfo{author}{\bibfnamefont{C.~S.} \bibnamefont{Gerving}},
  \bibinfo{author}{\bibfnamefont{T.~M.} \bibnamefont{Hoang}},
  \bibinfo{author}{\bibfnamefont{E.~M.} \bibnamefont{Bookjans}},
  \bibnamefont{and} \bibinfo{author}{\bibfnamefont{M.~S.}
  \bibnamefont{Chapman}}, \bibinfo{journal}{Nature Phys.}
  \textbf{\bibinfo{volume}{8}}, \bibinfo{pages}{305} (\bibinfo{year}{2012}).

\bibitem[{\citenamefont{Gross et~al.}(2011)\citenamefont{Gross, Strobel,
  Nicklas, Zibold, Bar-Gill, Kurizki, and Oberthaler}}]{Gross2011}
\bibinfo{author}{\bibfnamefont{C.}~\bibnamefont{Gross}},
  \bibinfo{author}{\bibfnamefont{H.}~\bibnamefont{Strobel}},
  \bibinfo{author}{\bibfnamefont{E.}~\bibnamefont{Nicklas}},
  \bibinfo{author}{\bibfnamefont{T.}~\bibnamefont{Zibold}},
  \bibinfo{author}{\bibfnamefont{N.}~\bibnamefont{Bar-Gill}},
  \bibinfo{author}{\bibfnamefont{G.}~\bibnamefont{Kurizki}}, \bibnamefont{and}
  \bibinfo{author}{\bibfnamefont{M.~K.} \bibnamefont{Oberthaler}},
  \bibinfo{journal}{Nature} \textbf{\bibinfo{volume}{480}},
  \bibinfo{pages}{219} (\bibinfo{year}{2011}).

\bibitem[{\citenamefont{L{\"u}cke et~al.}(2011)\citenamefont{L{\"u}cke,
  Scherer, Kruse, Pezze, Deuretzbacher, Hyllus, Topic, Peise, Ertmer, Arlt
  et~al.}}]{Lucke2011}
\bibinfo{author}{\bibfnamefont{B.}~\bibnamefont{L{\"u}cke}},
  \bibinfo{author}{\bibfnamefont{M.}~\bibnamefont{Scherer}},
  \bibinfo{author}{\bibfnamefont{J.}~\bibnamefont{Kruse}},
  \bibinfo{author}{\bibfnamefont{L.}~\bibnamefont{Pezze}},
  \bibinfo{author}{\bibfnamefont{F.}~\bibnamefont{Deuretzbacher}},
  \bibinfo{author}{\bibfnamefont{P.}~\bibnamefont{Hyllus}},
  \bibinfo{author}{\bibfnamefont{O.}~\bibnamefont{Topic}},
  \bibinfo{author}{\bibfnamefont{J.}~\bibnamefont{Peise}},
  \bibinfo{author}{\bibfnamefont{W.}~\bibnamefont{Ertmer}},
  \bibinfo{author}{\bibfnamefont{J.}~\bibnamefont{Arlt}}, \bibnamefont{et~al.},
  \bibinfo{journal}{Science} \textbf{\bibinfo{volume}{334}},
  \bibinfo{pages}{773} (\bibinfo{year}{2011}).

\bibitem[{\citenamefont{B{\"u}cker et~al.}(2011)\citenamefont{B{\"u}cker,
  Grond, Manz, Berrada, Betz, Koller, Hohenester, Schumm, Perrin, and
  Schmiedmayer}}]{Bucker:2011}
\bibinfo{author}{\bibfnamefont{R.}~\bibnamefont{B{\"u}cker}},
  \bibinfo{author}{\bibfnamefont{J.}~\bibnamefont{Grond}},
  \bibinfo{author}{\bibfnamefont{S.}~\bibnamefont{Manz}},
  \bibinfo{author}{\bibfnamefont{T.}~\bibnamefont{Berrada}},
  \bibinfo{author}{\bibfnamefont{T.}~\bibnamefont{Betz}},
  \bibinfo{author}{\bibfnamefont{C.}~\bibnamefont{Koller}},
  \bibinfo{author}{\bibfnamefont{U.}~\bibnamefont{Hohenester}},
  \bibinfo{author}{\bibfnamefont{T.}~\bibnamefont{Schumm}},
  \bibinfo{author}{\bibfnamefont{A.}~\bibnamefont{Perrin}}, \bibnamefont{and}
  \bibinfo{author}{\bibfnamefont{J.}~\bibnamefont{Schmiedmayer}},
  \bibinfo{journal}{Nature Phys.}  (\bibinfo{year}{2011}).

\bibitem[{\citenamefont{Est{\`e}ve et~al.}(2008)\citenamefont{Est{\`e}ve,
  Gross, Weller, Giovanazzi, and Oberthaler}}]{Esteve2008}
\bibinfo{author}{\bibfnamefont{J.}~\bibnamefont{Est{\`e}ve}},
  \bibinfo{author}{\bibfnamefont{C.}~\bibnamefont{Gross}},
  \bibinfo{author}{\bibfnamefont{A.}~\bibnamefont{Weller}},
  \bibinfo{author}{\bibfnamefont{S.}~\bibnamefont{Giovanazzi}},
  \bibnamefont{and} \bibinfo{author}{\bibfnamefont{M.~K.}
  \bibnamefont{Oberthaler}}, \bibinfo{journal}{Nature}
  \textbf{\bibinfo{volume}{455}}, \bibinfo{pages}{1216} (\bibinfo{year}{2008}).

\bibitem[{\citenamefont{Maussang et~al.}(2010)\citenamefont{Maussang, Marti,
  Schneider, Treutlein, Li, Sinatra, Long, {Est\`eve}, and
  Reichel}}]{Reichel:2010}
\bibinfo{author}{\bibfnamefont{K.}~\bibnamefont{Maussang}},
  \bibinfo{author}{\bibfnamefont{G.~E.} \bibnamefont{Marti}},
  \bibinfo{author}{\bibfnamefont{T.}~\bibnamefont{Schneider}},
  \bibinfo{author}{\bibfnamefont{P.}~\bibnamefont{Treutlein}},
  \bibinfo{author}{\bibfnamefont{Y.}~\bibnamefont{Li}},
  \bibinfo{author}{\bibfnamefont{A.}~\bibnamefont{Sinatra}},
  \bibinfo{author}{\bibfnamefont{R.}~\bibnamefont{Long}},
  \bibinfo{author}{\bibfnamefont{J.}~\bibnamefont{{Est\`eve}}},
  \bibnamefont{and} \bibinfo{author}{\bibfnamefont{J.}~\bibnamefont{Reichel}},
  \bibinfo{journal}{Phys. Rev. Lett.} \textbf{\bibinfo{volume}{105}},
  \bibinfo{pages}{080403} (\bibinfo{year}{2010}),
  \urlprefix\url{http://link.aps.org/doi/10.1103/PhysRevLett.105.080403}.

\bibitem[{\citenamefont{Kitagawa and Ueda}(1993{\natexlab{b}})}]{Ueda:1993}
\bibinfo{author}{\bibfnamefont{M.}~\bibnamefont{Kitagawa}} \bibnamefont{and}
  \bibinfo{author}{\bibfnamefont{M.}~\bibnamefont{Ueda}},
  \bibinfo{journal}{Phys. Rev. A} \textbf{\bibinfo{volume}{47}},
  \bibinfo{pages}{5138} (\bibinfo{year}{1993}{\natexlab{b}}),
  \urlprefix\url{http://link.aps.org/doi/10.1103/PhysRevA.47.5138}.

\bibitem[{\citenamefont{Sorensen et~al.}(2001)\citenamefont{Sorensen, Duan,
  Cirac, and Zoller}}]{Sorensen:2001}
\bibinfo{author}{\bibfnamefont{A.}~\bibnamefont{Sorensen}},
  \bibinfo{author}{\bibfnamefont{L.}~\bibnamefont{Duan}},
  \bibinfo{author}{\bibfnamefont{J.}~\bibnamefont{Cirac}}, \bibnamefont{and}
  \bibinfo{author}{\bibfnamefont{P.}~\bibnamefont{Zoller}},
  \bibinfo{journal}{Nature} \textbf{\bibinfo{volume}{409}}, \bibinfo{pages}{63}
  (\bibinfo{year}{2001}).

\bibitem[{\citenamefont{Li et~al.}(2009)\citenamefont{Li, Treutlein, Reichel,
  and Sinatra}}]{YunLi:2009}
\bibinfo{author}{\bibfnamefont{Y.}~\bibnamefont{Li}},
  \bibinfo{author}{\bibfnamefont{P.}~\bibnamefont{Treutlein}},
  \bibinfo{author}{\bibfnamefont{J.}~\bibnamefont{Reichel}}, \bibnamefont{and}
  \bibinfo{author}{\bibfnamefont{A.}~\bibnamefont{Sinatra}},
  \bibinfo{journal}{The European Physical Journal B}
  \textbf{\bibinfo{volume}{68}}, \bibinfo{pages}{365} (\bibinfo{year}{2009}),
  ISSN \bibinfo{issn}{1434-6028},
  \urlprefix\url{http://www.springerlink.com/index/10.1140/epjb/e2008-00472-6}.

\bibitem[{\citenamefont{Einstein et~al.}(1935)\citenamefont{Einstein, Podolsky,
  and Rosen}}]{EPR:1935}
\bibinfo{author}{\bibfnamefont{A.}~\bibnamefont{Einstein}},
  \bibinfo{author}{\bibfnamefont{B.}~\bibnamefont{Podolsky}}, \bibnamefont{and}
  \bibinfo{author}{\bibfnamefont{N.}~\bibnamefont{Rosen}},
  \bibinfo{journal}{Phys. Rev.} \textbf{\bibinfo{volume}{47}},
  \bibinfo{pages}{777} (\bibinfo{year}{1935}),
  \urlprefix\url{http://link.aps.org/doi/10.1103/PhysRev.47.777}.

\bibitem[{\citenamefont{He et~al.}(2011)\citenamefont{He, Reid, Vaughan, Gross,
  Oberthaler, and Drummond}}]{He:2011}
\bibinfo{author}{\bibfnamefont{Q.}~\bibnamefont{He}},
  \bibinfo{author}{\bibfnamefont{M.}~\bibnamefont{Reid}},
  \bibinfo{author}{\bibfnamefont{T.}~\bibnamefont{Vaughan}},
  \bibinfo{author}{\bibfnamefont{C.}~\bibnamefont{Gross}},
  \bibinfo{author}{\bibfnamefont{M.}~\bibnamefont{Oberthaler}},
  \bibnamefont{and} \bibinfo{author}{\bibfnamefont{P.}~\bibnamefont{Drummond}},
  \bibinfo{journal}{Phys. Rev. Lett.} \textbf{\bibinfo{volume}{106}},
  \bibinfo{pages}{120405} (\bibinfo{year}{2011}).

\bibitem[{\citenamefont{Bar-Gill et~al.}(2011)\citenamefont{Bar-Gill, Gross,
  Mazets, Oberthaler, and Kurizki}}]{BarGill:2011}
\bibinfo{author}{\bibfnamefont{N.}~\bibnamefont{Bar-Gill}},
  \bibinfo{author}{\bibfnamefont{C.}~\bibnamefont{Gross}},
  \bibinfo{author}{\bibfnamefont{I.}~\bibnamefont{Mazets}},
  \bibinfo{author}{\bibfnamefont{M.}~\bibnamefont{Oberthaler}},
  \bibnamefont{and} \bibinfo{author}{\bibfnamefont{G.}~\bibnamefont{Kurizki}},
  \bibinfo{journal}{Phys. Rev. Lett.} \textbf{\bibinfo{volume}{106}},
  \bibinfo{pages}{120404} (\bibinfo{year}{2011}).

\bibitem[{\citenamefont{Julsgaard et~al.}(2001)\citenamefont{Julsgaard,
  Kozhekin, and Polzik}}]{Julsgaard2001}
\bibinfo{author}{\bibfnamefont{B.}~\bibnamefont{Julsgaard}},
  \bibinfo{author}{\bibfnamefont{a.}~\bibnamefont{Kozhekin}}, \bibnamefont{and}
  \bibinfo{author}{\bibfnamefont{E.~S.} \bibnamefont{Polzik}},
  \bibinfo{journal}{Nature} \textbf{\bibinfo{volume}{413}},
  \bibinfo{pages}{400} (\bibinfo{year}{2001}), ISSN \bibinfo{issn}{0028-0836},
  \urlprefix\url{http://www.ncbi.nlm.nih.gov/pubmed/11574882}.

\bibitem[{\citenamefont{Wiseman et~al.}(2007)\citenamefont{Wiseman, Jones, and
  Doherty}}]{Wiseman2007}
\bibinfo{author}{\bibfnamefont{H.~M.} \bibnamefont{Wiseman}},
  \bibinfo{author}{\bibfnamefont{S.~J.} \bibnamefont{Jones}}, \bibnamefont{and}
  \bibinfo{author}{\bibfnamefont{A.~C.} \bibnamefont{Doherty}},
  \bibinfo{journal}{Physical Review Letters} \textbf{\bibinfo{volume}{98}},
  \bibinfo{pages}{140402} (\bibinfo{year}{2007}), ISSN
  \bibinfo{issn}{0031-9007},
  \urlprefix\url{http://link.aps.org/doi/10.1103/PhysRevLett.98.140402}.

\bibitem[{\citenamefont{Opanchuk et~al.}(2012)\citenamefont{Opanchuk, He, Reid,
  and Drummond}}]{Opanchuk2012}
\bibinfo{author}{\bibfnamefont{B.}~\bibnamefont{Opanchuk}},
  \bibinfo{author}{\bibfnamefont{Q.~Y.} \bibnamefont{He}},
  \bibinfo{author}{\bibfnamefont{M.~D.} \bibnamefont{Reid}}, \bibnamefont{and}
  \bibinfo{author}{\bibfnamefont{P.~D.} \bibnamefont{Drummond}},
  \bibinfo{journal}{Physical Review A} \textbf{\bibinfo{volume}{86}},
  \bibinfo{pages}{023625} (\bibinfo{year}{2012}), ISSN
  \bibinfo{issn}{1050-2947},
  \urlprefix\url{http://link.aps.org/doi/10.1103/PhysRevA.86.023625}.

\bibitem[{\citenamefont{Briegel et~al.}(2000)\citenamefont{Briegel, Calarco,
  Jaksch, Cirac, and Zoller}}]{Zoller:2000}
\bibinfo{author}{\bibfnamefont{H.}~\bibnamefont{Briegel}},
  \bibinfo{author}{\bibfnamefont{T.}~\bibnamefont{Calarco}},
  \bibinfo{author}{\bibfnamefont{D.}~\bibnamefont{Jaksch}},
  \bibinfo{author}{\bibfnamefont{J.}~\bibnamefont{Cirac}}, \bibnamefont{and}
  \bibinfo{author}{\bibfnamefont{P.}~\bibnamefont{Zoller}},
  \bibinfo{journal}{Journal of Modern Optics} \textbf{\bibinfo{volume}{47}},
  \bibinfo{pages}{415} (\bibinfo{year}{2000}).

\bibitem[{\citenamefont{Treutlein et~al.}(2006)\citenamefont{Treutlein,
  H\"{a}nsch, Reichel, Negretti, Cirone, and Calarco}}]{Treutlein2006}
\bibinfo{author}{\bibfnamefont{P.}~\bibnamefont{Treutlein}},
  \bibinfo{author}{\bibfnamefont{T.}~\bibnamefont{H\"{a}nsch}},
  \bibinfo{author}{\bibfnamefont{J.}~\bibnamefont{Reichel}},
  \bibinfo{author}{\bibfnamefont{A.}~\bibnamefont{Negretti}},
  \bibinfo{author}{\bibfnamefont{M.}~\bibnamefont{Cirone}}, \bibnamefont{and}
  \bibinfo{author}{\bibfnamefont{T.}~\bibnamefont{Calarco}},
  \bibinfo{journal}{Physical Review A} \textbf{\bibinfo{volume}{74}},
  \bibinfo{pages}{022312} (\bibinfo{year}{2006}), ISSN
  \bibinfo{issn}{1050-2947},
  \urlprefix\url{http://link.aps.org/doi/10.1103/PhysRevA.74.022312}.

\bibitem[{\citenamefont{Mandel et~al.}(2003)\citenamefont{Mandel, Greiner,
  Widera, Rom, H\"ansch, and Bloch}}]{Bloch:2003}
\bibinfo{author}{\bibfnamefont{O.}~\bibnamefont{Mandel}},
  \bibinfo{author}{\bibfnamefont{M.}~\bibnamefont{Greiner}},
  \bibinfo{author}{\bibfnamefont{A.}~\bibnamefont{Widera}},
  \bibinfo{author}{\bibfnamefont{T.}~\bibnamefont{Rom}},
  \bibinfo{author}{\bibfnamefont{T.}~\bibnamefont{H\"ansch}}, \bibnamefont{and}
  \bibinfo{author}{\bibfnamefont{I.}~\bibnamefont{Bloch}},
  \bibinfo{journal}{Nature} \textbf{\bibinfo{volume}{425}},
  \bibinfo{pages}{937} (\bibinfo{year}{2003}).

\bibitem[{\citenamefont{P. et~al.}(2009)\citenamefont{P., M.F., J., Reichel,
  T.W., and P.}}]{Boehi:2009}
\bibinfo{author}{\bibfnamefont{P.}~\bibnamefont{B\"ohi}},
  \bibinfo{author}{\bibfnamefont{M.F.}~\bibnamefont{Riedler}},
  \bibinfo{author}{\bibfnamefont{J.}~\bibnamefont{Hoffrogge}},
  \bibinfo{author}{\bibfnamefont{J.}~\bibnamefont{Reichel}},
  \bibinfo{author}{\bibfnamefont{T.W.}~\bibnamefont{H\"ansch}}, \bibnamefont{and}
  \bibinfo{author}{\bibfnamefont{P.}~\bibnamefont{Treutlein}},
  \bibinfo{journal}{Nature Physics} \textbf{\bibinfo{volume}{5}},
  \bibinfo{pages}{592} (\bibinfo{year}{2009}).

\bibitem[{\citenamefont{Castin}(2001)}]{LesHouches99}
\bibinfo{author}{\bibfnamefont{Y.}~\bibnamefont{Castin}},
  \emph{\bibinfo{title}{Bose-Einstein Condensates in Atomic Gases, Lecture
  notes of 1999 Les Houches summer school}} (\bibinfo{publisher}{EDP Sciences
  and Springer-Verlag}, \bibinfo{address}{Les Ulis/Berlin},
  \bibinfo{year}{2001}).

\bibitem[{\citenamefont{Ferrini et~al.}(2010)\citenamefont{Ferrini, Spehner,
  Minguzzi, and Hekking}}]{ferrini2010a}
\bibinfo{author}{\bibfnamefont{G.}~\bibnamefont{Ferrini}},
  \bibinfo{author}{\bibfnamefont{D.}~\bibnamefont{Spehner}},
  \bibinfo{author}{\bibfnamefont{A.}~\bibnamefont{Minguzzi}}, \bibnamefont{and}
  \bibinfo{author}{\bibfnamefont{F.~W.~J.} \bibnamefont{Hekking}},
  \bibinfo{journal}{Phys. Rev. A} \textbf{\bibinfo{volume}{82}},
  \bibinfo{pages}{033621} (\bibinfo{year}{2010}).


\bibitem[{\citenamefont{Yurke and Stoler}(1986)}]{yurke1986}
\bibinfo{author}{\bibfnamefont{B.}~\bibnamefont{Yurke}} \bibnamefont{and}
  \bibinfo{author}{\bibfnamefont{D.}~\bibnamefont{Stoler}},
  \bibinfo{journal}{Phys. Rev. Lett.} \textbf{\bibinfo{volume}{57}},
  \bibinfo{pages}{13} (\bibinfo{year}{1986}),
  \urlprefix\url{http://link.aps.org/doi/10.1103/PhysRevLett.57.13}.

\bibitem[{\citenamefont{Stoler}(1971)}]{stoler1971}
\bibinfo{author}{\bibfnamefont{D.}~\bibnamefont{Stoler}},
  \bibinfo{journal}{Phys. Rev. D} \textbf{\bibinfo{volume}{4}},
  \bibinfo{pages}{2309} (\bibinfo{year}{1971}),
  \urlprefix\url{http://link.aps.org/doi/10.1103/PhysRevD.4.2309}.

\bibitem[{\citenamefont{Erhard et~al.}(2004)\citenamefont{Erhard, Schmaljohann,
  Kronj\"ager, Bongs, and Sengstock}}]{Sengstock:2004}
\bibinfo{author}{\bibfnamefont{M.}~\bibnamefont{Erhard}},
  \bibinfo{author}{\bibfnamefont{H.}~\bibnamefont{Schmaljohann}},
  \bibinfo{author}{\bibfnamefont{J.}~\bibnamefont{Kronj\"ager}},
  \bibinfo{author}{\bibfnamefont{K.}~\bibnamefont{Bongs}}, \bibnamefont{and}
  \bibinfo{author}{\bibfnamefont{K.}~\bibnamefont{Sengstock}},
  \bibinfo{journal}{Phys. Rev. A} \textbf{\bibinfo{volume}{69}},
  \bibinfo{pages}{032705} (\bibinfo{year}{2004}),
  \urlprefix\url{http://link.aps.org/doi/10.1103/PhysRevA.69.032705}.

\bibitem[{\citenamefont{Sinatra and Castin}(1998)}]{Sinatra:1998}
\bibinfo{author}{\bibfnamefont{A.}~\bibnamefont{Sinatra}} \bibnamefont{and}
  \bibinfo{author}{\bibfnamefont{Y.}~\bibnamefont{Castin}},
  \bibinfo{journal}{Eur. Phys. Jour. B} \textbf{\bibinfo{volume}{4}},
  \bibinfo{pages}{247} (\bibinfo{year}{1998}).

\bibitem[{\citenamefont{Pawlowski et~al.}(2013)\citenamefont{Pawlowski,
  Spehner, Minguzzi, and Ferrini}}]{Pawlowski:2013}
\bibinfo{author}{\bibfnamefont{K.}~\bibnamefont{Pawlowski}},
  \bibinfo{author}{\bibfnamefont{D.}~\bibnamefont{Spehner}},
  \bibinfo{author}{\bibfnamefont{A.}~\bibnamefont{Minguzzi}}, \bibnamefont{and}
  \bibinfo{author}{\bibfnamefont{G.}~\bibnamefont{Ferrini}},
  \bibinfo{journal}{Phys. Rev. A} \textbf{\bibinfo{volume}{88}},
  \bibinfo{pages}{013606} (\bibinfo{year}{2013}),
  \urlprefix\url{http://link.aps.org/doi/10.1103/PhysRevA.88.013606}.

\bibitem[{\citenamefont{Walls and Milburn}(1994)}]{walls_milburn_book_94}
\bibinfo{author}{\bibfnamefont{D.}~\bibnamefont{Walls}} \bibnamefont{and}
  \bibinfo{author}{\bibfnamefont{G.}~\bibnamefont{Milburn}},
  \emph{\bibinfo{title}{Quantum Optics}} (\bibinfo{publisher}{Springer-Verlag},
  \bibinfo{address}{Berlin Heidelberg New York}, \bibinfo{year}{1994}).

\bibitem[{\citenamefont{Cavalcanti et~al.}(2011)\citenamefont{Cavalcanti, He,
  Reid, and Wiseman}}]{Cavalcanti2011}
\bibinfo{author}{\bibfnamefont{E.~G.} \bibnamefont{Cavalcanti}},
  \bibinfo{author}{\bibfnamefont{Q.~Y.} \bibnamefont{He}},
  \bibinfo{author}{\bibfnamefont{M.~D.} \bibnamefont{Reid}}, \bibnamefont{and}
  \bibinfo{author}{\bibfnamefont{H.~M.} \bibnamefont{Wiseman}},
  \bibinfo{journal}{Physical Review A} \textbf{\bibinfo{volume}{032115}},
  \bibinfo{pages}{1} (\bibinfo{year}{2011}).

\bibitem[{\citenamefont{Sinatra et~al.}(2012)\citenamefont{Sinatra,
  Dornstetter, and Castin}}]{Frontiers:2012}
\bibinfo{author}{\bibfnamefont{A.}~\bibnamefont{Sinatra}},
  \bibinfo{author}{\bibfnamefont{J.-C.} \bibnamefont{Dornstetter}},
  \bibnamefont{and} \bibinfo{author}{\bibfnamefont{Y.}~\bibnamefont{Castin}},
  \bibinfo{journal}{Front. Phys.} \textbf{\bibinfo{volume}{7}},
  \bibinfo{pages}{86} (\bibinfo{year}{2012}).

\bibitem[{\citenamefont{Ockeloen et~al.}(2013)\citenamefont{Ockeloen, Schmied,
  Riedel, and Treutlein}}]{Ockeloen:2013}
\bibinfo{author}{\bibfnamefont{C.~F.} \bibnamefont{Ockeloen}},
  \bibinfo{author}{\bibfnamefont{R.}~\bibnamefont{Schmied}},
  \bibinfo{author}{\bibfnamefont{M.~F.} \bibnamefont{Riedel}},
  \bibnamefont{and}
  \bibinfo{author}{\bibfnamefont{P.}~\bibnamefont{Treutlein}},
  \bibinfo{journal}{arXiv:1303.1313}  (\bibinfo{year}{2013}).

\bibitem[{\citenamefont{Wineland et~al.}(1994)\citenamefont{Wineland, Bollinger, Itano, and Heinzen}}]{Wineland:1994}
\bibinfo{author}{\bibfnamefont{D.J.}~\bibnamefont{Wineland}},
  \bibinfo{author}{\bibfnamefont{J.J.}~\bibnamefont{Bollinger}}, 
  \bibinfo{author}{\bibfnamefont{W.M.}~\bibnamefont{Itano}}, \bibnamefont{and}
  \bibinfo{author}{\bibfnamefont{D.J.}~\bibnamefont{Heinzen}},
  \bibinfo{journal}{Phys. Rev. A} \textbf{\bibinfo{volume}{50}},
  \bibinfo{pages}{67} (\bibinfo{year}{1994}).

\bibitem[{\citenamefont{Li et~al.}(2008)\citenamefont{Li, Castin, and
  Sinatra}}]{LiYun:2008}
\bibinfo{author}{\bibfnamefont{Y.}~\bibnamefont{Li}},
  \bibinfo{author}{\bibfnamefont{Y.}~\bibnamefont{Castin}}, \bibnamefont{and}
  \bibinfo{author}{\bibfnamefont{A.}~\bibnamefont{Sinatra}},
  \bibinfo{journal}{Phys. Rev. Lett.} \textbf{\bibinfo{volume}{100}},
  \bibinfo{pages}{210401} (\bibinfo{year}{2008}).

\bibitem[{\citenamefont{Byrnes}(2013)}]{Byrnes:2013}
\bibinfo{author}{\bibfnamefont{T.}~\bibnamefont{Byrnes}},
  \bibinfo{journal}{Phys. Rev. A}  \textbf{\bibinfo{volume}{88}},
  \bibinfo{pages}{023609} (\bibinfo{year}{2013}).


\end{thebibliography}

\end{document}